\documentclass[12pt]{article}
\pdfoutput=1
 \usepackage{geometry}
 
\usepackage[nosort]{cite}

\usepackage[dvipsnames]{xcolor}
\usepackage{cancel}

 \usepackage{caption}
\usepackage{epsfig}
\usepackage{amsfonts}
\usepackage{amscd}
\usepackage{latexsym}
\usepackage{dsfont}
\usepackage{amsmath,amssymb}
\usepackage{authblk}
\usepackage{verbatim}
\usepackage{setspace}
\usepackage{color}
\usepackage{fancyhdr}
\usepackage{cite}
\usepackage{hyperref}
\usepackage{tikz}
\usepackage[pdf]{pstricks}
\usepackage{subcaption}
\usepackage{slashed}
\usepackage{multirow}
\usetikzlibrary{calc}
\usetikzlibrary{topaths}
\usetikzlibrary{decorations}
\usetikzlibrary{decorations.pathmorphing}
\usetikzlibrary{arrows,decorations.markings,cd}
\usetikzlibrary{calc,arrows,cd,decorations.markings,snakes}
\usetikzlibrary{arrows.meta}
\tikzset{
->-/.style args={#1rotate#2}{decoration={markings, mark=at position #1 with {\arrow[scale=1.5,rotate = #2 ]{stealth}}}, postaction={decorate}}
}
\usetikzlibrary{shapes.geometric}
\usetikzlibrary{knots}
\usepackage{tikz-3dplot}

 \usepackage{draft}
\usepackage{hyperref}
\usepackage{cleveref}
\usepackage{graphicx}
\usepackage{cite}
\usepackage{mciteplus}
\usepackage{skak}
\usepackage{bbm}
\usepackage{ulem}
 
\usepackage[bottom,flushmargin]{footmisc}

\numberwithin{equation}{section}

\def\ie{\begin{equation}\begin{aligned}}
\def\fe{\end{aligned}\end{equation}}

\renewcommand{\H}{\mathcal{H}}
\renewcommand{\A}{\mathcal{A}}
\renewcommand{\B}{\mathcal{B}}
\newcommand{\R}{\mathcal{R}}

\renewcommand{\L}{\mathcal{L}}

\begin{document}

\begin{titlepage}
\hfill  MIT-CTP/5853, YITP-SB-2025-06

\title{Additivity, Haag duality, and non-invertible symmetries}

\author{Shu-Heng Shao$^{1,2}$, Jonathan Sorce$^{1}$, and Manu Srivastava$^{1}$} 

\textit{${}^{1}$Center for Theoretical Physics - a Leinweber Institute, Massachusetts Institute of Technology, Cambridge, MA}

\vspace{2pt}
\textit{${}^{2}$Yang Institute for Theoretical Physics, Stony Brook University, Stony Brook, NY}

 \vspace{20pt}
\abstract{The algebraic approach to quantum field theory focuses on the properties of local algebras, whereas the study of (possibly non-invertible) global symmetries emphasizes global aspects of the theory and spacetime. We study connections between these two perspectives by examining how either of two core algebraic properties --- ``additivity'' or ``Haag duality'' --- is violated in a 1+1D CFT or lattice model restricted to the symmetric sector of a general  global symmetry. 
For the Verlinde symmetry of a bosonic diagonal RCFT, we find that additivity is violated whenever the symmetry algebra contains an invertible element, while Haag duality is violated whenever it contains a non-invertible element. 
We find similar phenomena for the Kramers-Wannier and Rep(D$_8$)  non-invertible symmetries on spin chains.
}

\end{titlepage}

\tableofcontents

\section{Introduction}
 
Quantum field theory (QFT) is a famously complex subject, and many different perspectives have been used in attempts to achieve a unified understanding of its features.
Two perspectives that have become popular are to study QFT using generalized symmetries \cite{Gaiotto:2014kfa}, or to study QFT using the structure of local algebras \cite{Haag:axioms, Haag:book}. 
Relatively few connections have been made between these two communities, though inroads have been made in \cite{Harlow:2018tng,Casini:2020rgj, Casini:completeness, Benedetti:Noether,Benedetti:2022ofj,Witten:2023qsv,Fliss:2023dze,Benedetti:2023owa,Benedetti:2024dku,Benedetti:2024utz}.
The goal of the present paper is to strengthen the connection between these two ways of thinking about QFT by studying connections between the algebraic approach and non-invertible symmetries.
 
As we explain further in section \ref{sec:algebras}, the algebraic approach to studying quantum theory involves a ``net of local algebras'', which is a way of assigning an algebra $\A(\R)$ to each region $\R$.
The properties of this assignment encode deep structural facts about the theory one is studying; for example, in \cite{DHR-1, DHR-2}, the structure of this assignment in a gauge theory was shown to be related to the structure of superselection sectors. More recently, in \cite{Benedetti:2024dku}, based on earlier work in \cite{Kawahigashi:modular, Rehren:modular}, the structure of the assignment in a 1+1D conformal field theory was shown to be related to the modular invariance of the theory's torus partition function. 
On the other hand, as reviewed in \cite{McGreevy:2022oyu,Cordova:2022ruw,Schafer-Nameki:2023jdn,Brennan:2023mmt,Bhardwaj:2023kri,Shao:2023gho}, generalized symmetries consist of invertible or non-invertible topological defects that can be used to constrain the correlation functions of a quantum field theory, leading to novel selection rules and many other interesting phenomena.

The chief difficulty in connecting the algebraic and symmetry perspectives is that the algebraic framework was initially designed to describe the physics of local fields, while the symmetry perspective emphasizes the importance of extended operators. 
This has caused a great deal of confusion in determining what the ``correct'' way to assign algebras to regions in the presence of extended operators is, with different research groups having different opinions on how this question should be answered. 
We do not aim to resolve this debate, but we do hope to introduce new connections between the algebraic literature and the literature on non-invertible global symmetries.

Concretely, in this work we will study finite global symmetry algebras in 1+1D systems that will generally have both invertible and non-invertible elements.
Restricting to the sector of the 1+1D theory that is invariant under this global symmetry algebra gives rise to a new ``theory'' that may be pathological. 
For example, it is well known that while the $\mathbb{Z}_2$-even sector of the 1+1D Ising CFT is closed under operator product expansions, it is not modular invariant and does not have well-defined partition functions on general manifolds. 
While modular invariance might be a natural rule to impose in relativistic QFT, it is unclear what it means in more general setups such as in lattice models. 
Indeed, there have been many recent discussions of sectors of full-fledged lattice systems under global symmetry constraints in the study of topological orders and dualities  \cite{Ji:2019jhk,Kong:2020cie,Wu:2020yxa,Ji:2021esj,Chatterjee:2022kxb,Liu:2022cxc,Chatterjee:2022jll,Jones:2023ptg,Jones:2023xew,Inamura:2023ldn,Jones:2024lws}.

Interestingly, as pointed out in \cite{Haag:book}, the local algebras of  $\mathbb{Z}_2$-even sector  have an unusual structure in that they cannot simultaneously obey two core properties in algebraic QFT, known as \textit{Haag duality} and \textit{additivity}. 
This observation motivates us to study how the properties of additivity and Haag duality, which characterize the structure of the local algebras in a quantum theory, are preserved or violated under restriction to a symmetric sector of a finite symmetry algebra. 
In the examples we analyze, our general observation will be that for a natural assignment of algebras that we define in section \ref{sec:invertible}, the property of additivity is violated when the symmetry algebra contains an invertible element, while the property of Haag duality is violated when the symmetry algebra contains a non-invertible element.
We will show this both in the continuum setting of rational conformal field theories with (generally non-invertible) Verlinde lines, and in two explicit lattice models containing non-invertible duality defects.
This observation connects the additivity and Haag duality properties that have been emphasized in \cite{Casini:2020rgj, Casini:completeness, Benedetti:Noether, Benedetti:2024dku} with the non-invertible defects reviewed in \cite{Bhardwaj:2017xup, Chang:2018iay}.  

The outline of the paper is as follows.
In section \ref{sec:algebras}, we review the algebraic framework for studying local quantum theories, and explain the ``additivity'' and ``Haag duality'' properties that will be studied in later sections.
In section \ref{sec:invertible}, we study the example of the $\mathbb{Z}_2$-even sector of the Ising model in more detail, and explain how a natural choice of algebraic assignment violates additivity but not Haag duality.
In section \ref{sec:continuum}, we study diagonal rational conformal field theories and the restriction to the symmetric sector of the finite symmetry generated by Verlinde lines; we show that additivity is violated if there is an invertible line, while Haag duality is violated if there is a non-invertible line.
In section \ref{sec:lattice}, we study similar phenomena in two examples of lattice systems with non-invertible symmetries.

While not explicitly overlapping with the present work, we wish to emphasize that there is a wealth of interesting literature on symmetry in algebraic quantum field theory that may be relevant to the modern paradigm of studying quantum theories using generalized symmetries. 
For an incomplete list of references, see \cite{DHR-1, DHR-2, Longo:1989tt, Longo:1990zp, Kawahigashi:modular, Muger:2000rc, Rehren:modular, Carpi:VOA, Kawahigashi:2D, Evans:subfactors, Jones:2023xew}.

\section{Algebraic notions}
 \label{sec:algebras}

In the algebraic approach to quantum theory developed in \cite{Haag:axioms, Haag:book}, subsystems of a quantum system are described by algebras of operators represented on a common Hilbert space $\H$.
The states in $\H$ are the global states of the full system; for any state $|\psi \rangle \in \H$ and any algebra $\A$ defining a subsystem, the set of expectation values
\begin{equation}
    \langle \psi | a | \psi \rangle, \qquad a \in \A
\end{equation}
characterizes the restriction of $|\psi\rangle$ to the subsystem described by $\A$.

The algebras describing subsystems are typically taken to be \textit{von Neumann algebras}.
These are algebras of bounded operators that are closed under adjoints, and that are complete with respect to a naturally physical topology.
For precise definitions, see the physicist-oriented reviews \cite{Witten:notes, Sorce:notes}.
A useful equivalent definition is that a von Neumann algebra is an algebra of bounded operators that is equal to its own \textit{double commutant}.
If we denote by $\B(\H)$ the space of all bounded operators on $\H,$ then the commutant of $\A$ is the set
\begin{equation}
    \A' \equiv \{ a' \in \B(\H)\, | \, [a', a] = 0 \text{ for all } a \in \A\},
\end{equation}
and $\A$ is a von Neumann algebra if and only if it satisfies $\A = (\A')',$ often written more succinctly as $\A = \A''.$

A quantum system typically comes with some notion of locality in that the degrees of freedom can be partitioned into distinct regions.
In a relativistic quantum field theory, for example, the degrees of freedom are naturally partitioned among causally complete regions of spacetime.
In a lattice system, the degrees of freedom are naturally partitioned among subsets of the lattice sites.
In any case, a local quantum theory contains a set of regions that contain subsystems, and to study these subsystems algebraically, one assigns a von Neumann algebra $\A(\R)$ to each region $\R$.
On physical grounds, these algebras should satisfy some natural properties discussed at length in \cite{Haag:axioms, Haag:book}.
The most basic of these is ``isotony,'' which is just the statement
\begin{equation}
    \R_1 \subseteq \R_2 \quad \Rightarrow \quad \A(\R_1) \subseteq \A(\R_2).
\end{equation}
Another natural condition is that there should be a notion of a complementary region $\R'$ for each region $\R$, and that by locality, the algebras should satisfy the restriction 
\begin{equation}\label{causal}
    \A(\R') \subseteq \A(\R)',
\end{equation}
since everything in $\R'$ should commute with everything in $\R.$
In a relativistic QFT, $\R'$ is the causal complement of $\R$, while on the lattice, $\R'$ is the complementary subset of lattice sites.

In the strict algebraic approach to quantum physics advocated in \cite{Haag:axioms, Haag:book}, the assignment of algebras to regions is part of the definition of a quantum theory.
However, if one has an independent definition of the theory --- for example a lattice Hamiltonian or a field theory Lagrangian --- then there is typically some freedom in how one defines the map $\R \mapsto \A(\R).$
There may be a ``correct'' assignment that is the most informative in terms of understanding the theory, but one is free to study other assignments and ask about their properties.
A typical example, which was studied in \cite{Casini:2020rgj, Casini:completeness}, is a theory with a higher-form global symmetry \cite{Gaiotto:2014kfa}.
Such a theory contains extended operators that cannot be divided up into a product of local pieces.
In such cases, one must decide whether such operators are included in the ``local algebras'' $\A(\R)$ or not.
Our preference is to include these operators in the local algebras, but this is not a universal point of view --- the opposite choice is made in \cite{Casini:2020rgj, Casini:completeness, Benedetti:2024dku}.
So in the present work, we will remain agnostic about what the ``correct'' choice of assignment of algebras to regions is, and simply study the properties of an assignment that we find natural.
Furthermore, we will only discuss quantum systems without higher-form global symmetries in this paper, leaving a detailed exposition of the connection to higher-form symmetries to a companion paper \cite{paper2}.

The specific properties that we will study are \textit{additivity} and \textit{Haag duality}. 
Additivity is the property that the algebra associated to a region is generated by the algebras associated to subregions.
For two algebras $\A(\R_1)$ and $\A(\R_2)$, we define $\A(\R_1) \vee \A(\R_2)$ to be the smallest von Neumann algebra containing both $\A(\R_1)$ and $\A(\R_2)$; an equivalent definition is
\begin{equation}
    \A(\R_1) \vee \A(\R_2) \equiv (\A(\R_1) \cup \A(\R_2))''.
\end{equation}
An assignment of algebras to regions is said to satisfy \textit{additivity} if we have
\begin{equation}\label{additivity}
    \A(\R_1 \cup \R_2)
        = \A(\R_1) \vee \A(\R_2).
\end{equation}

Haag duality is the property that the algebra associated to a region is ``maximal'' with respect to the algebra associated with the complementary region.
Concretely, we explained above that locality imposes the requirement \eqref{causal}.
An assignment of algebras to regions is said to satisfy \textit{Haag duality} if the inclusion is saturated, so that we always have
\begin{equation}
    \A(\R') = \A(\R)'
\end{equation}
for every region $\R$.
In other words,  Haag duality states that the algebra of the complement is the commutant of the algebra, which can be viewed as a complement/commutant duality.\footnote{The word ``duality" is also used in many other contexts in theoretical physics for different notions. See, for example, \cite{Pace:2024oys} for a recent survey.  }

\section{Additivity violation and invertible global symmetry} \label{sec:invertible}

As discussed in the introduction and in section \ref{sec:algebras}, the properties of additivity and Haag duality may be violated when one restricts to the symmetric sector of a global symmetry.
In this section, we will illustrate this using a simple 1+1D lattice model, which will allow us to clarify our general convention for the assignment of algebras to local regions, as well as the basic mechanism by which additivity is violated.
Since our focus is solely on the operator algebra, we do not need to specify a particular Hamiltonian.
However, one may keep the transverse-field Ising Hamiltonian in mind as a concrete example. 

The space of our model is a periodic spin chain with $L$ links. 
We place a qubit on every link and label the links by $\ell=1,2,\cdots, L$ with $\ell\sim \ell+L$. 
The total Hilbert space ${\cal H}=\bigotimes_{\ell=1}^L {\cal H}_\ell$ is $2^L$-dimensional and  is a tensor product of local qubit Hilbert spaces ${\cal H}_\ell\simeq \mathbb{C}^2$. 
The complete operator algebra ${\cal A}(S^1)$ is just the algebra of $2^L\times 2^L$ matrices
\ie\label{matrix}
{\cal A}(S^1) = \text{Mat}(2^L, \mathbb{C})\,,
\fe 
where $S^1$ denotes the entire space. 
Given a region $\R$ (which is generally a disjoint union of intervals), there is a natural assignment of an algebra $\A(\R)$, which is just the matrix algebra acting on the Hilbert space factor $\bigotimes_{\ell \in \R} \H_{\ell}$.
This assignment clearly satisfies both additivity and Haag duality. 

This lattice algebra contains a global $\mathbb{Z}_{2}$ operator
\ie
U=  \prod_{\ell=1}^L X_\ell \,,~~~~U^2=1\,,
\fe 
where $X_\ell,Z_\ell$ are the Pauli operators acting on the qubit of the $\ell$-th link.
When we restrict to the $\mathbb{Z}_2$-even sector, we restrict to the subalgebra ${\cal A}_{\mathbb{Z}_2}(S^1) \subseteq {\cal A}(S^1),$ which consists of operators that commute with $U$.  
In the Ising model, ${\cal A}_{\mathbb{Z}_2}(S^1)$ is known as the bond algebra \cite{2009PhRvB..79u4440N,Cobanera:2009as,Cobanera:2011wn} for the $\mathbb{Z}_2$ operator $U$, which in this case is 
\ie
{\cal A}_{\mathbb{Z}_2}(S^1)
&\equiv
\Big\{ 
{\cal O}\in \text{Mat}(2^L ,\mathbb{C}) ~\Big|  {\cal O} U =U{\cal O}
\Big\}\\
&=  \Biggl\langle
~X_\ell~,~ Z_{\ell}Z_{\ell+1}~
 \Biggr\rangle_{\ell=1,2,\cdots ,L} \,.
\fe
This subalgebra acts irreducibly on the $\mathbb{Z}_2$-even subspace of ${\cal H}$, defined as 
\begin{equation}
    {\cal H}_{\mathbb{Z}_2} = \left\{ \ket{\phi} \in {\cal H} ~| ~U\ket{\phi}=\ket{\phi}\right\},
\end{equation} which is no longer a tensor product Hilbert space. This subalgebra has been emphasized in \cite{Ji:2019jhk} in the context of topological order and dualities.

To study the properties of additivity and Haag duality, we need to define an assignment of algebras ${\cal A}_{\mathbb{Z}_2}(\R) \subseteq \A_{\mathbb{Z}_2}(S^1)$ for general regions $\mathcal{R}$.
As discussed in section \ref{sec:algebras}, there is no canonical choice of this assignment. 
For example, the operator $U$ belongs to $\A_{\mathbb{Z}_2}(S^1),$ but it is up to us to decide whether it belongs to ${\cal A}_{\mathbb{Z}_2}(\R)$ if $\R$ is not the entire circle.
Naively, because $U$ has support on every link, we might want to say that $U$ is not in ${\cal A}_{\mathbb{Z}_2}(\R)$. 
On the other hand, the operator $U$ acts trivially on ${\cal H}_{\mathbb{Z}_2}$, so it should be identified with the identity operator, which belongs to the algebra of every region $\R$.

In what follows, we will adopt the following physically motivated definition of $\A(\R)$.
We define an operator to belong to $\mathcal{A}_{\mathbb{Z}_2}(\R)$ if the relation $U = 1 $ allows us to express the operator in a form that is entirely supported within $\R$. 
Equivalently, we say an operator $\cal O$ belongs to $\mathcal{A}_{\mathbb{Z}_2}(\R)$ if there exists an operator in $\A(\R)$ that acts the same way as $\cal O$ on the subspace $\H_{\mathbb{Z}_2}.$
With this choice of assignment, $U$ belongs to ${\cal A}_{\mathbb{Z}_2}(\R)$ for every choice of $\R$.\footnote{See \cite{Fliss:2023dze} for similar algebra assignments in topological field theories.}

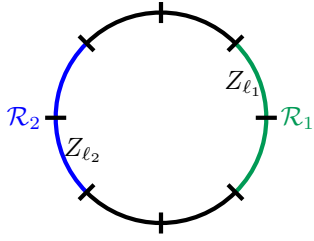
\begin{figure}
\centering
\begin{tikzpicture}[scale=0.7]  
    \draw[ultra thick] (45:2) arc[start angle=45, end angle=135, radius=2];

      \draw[ultra thick] (225:2) arc[start angle=225, end angle=315, radius=2];

    \draw[ultra thick, ForestGreen] (315:2) arc[start angle=-45, end angle=45, radius=2];

    \draw[ultra thick, blue] (135:2) arc[start angle=135, end angle=225, radius=2];

    \foreach \angle in {0, 45, 90, 135, 180, 225, 270, 315} {
      
        \pgfmathsetmacro\x{2*cos(\angle)}
        \pgfmathsetmacro\y{2*sin(\angle)}
        
        \pgfmathsetmacro\dx{0.2*cos(\angle)}
        \pgfmathsetmacro\dy{0.2*sin(\angle)}
        \draw[ultra thick] (\x-\dx, \y-\dy) -- (\x+\dx, \y+\dy);
    }

    \pgfmathsetmacro\x{1.7*cos(22.5)}
    \pgfmathsetmacro\y{1.7*sin(22.5)}
    \node at (\x, \y) {\footnotesize \textbf{$Z_{\ell_1}$}};

    \pgfmathsetmacro\x{1.6*cos(202.5)}
    \pgfmathsetmacro\y{1.6*sin(202.5)}
    \node at (\x, \y) {\footnotesize \textbf{$Z_{\ell_2}$}};

    \node at (2.6, 0) {\footnotesize \textbf{\color{ForestGreen}$\R_1$}};
    \node at (-2.6, 0) {\footnotesize \textbf{\color{blue}$\R_2$}};

\end{tikzpicture}
\caption{Additivity is violated by a pair of $\mathbb{Z}_2$-odd operators.}\label{fig:additivity}
\end{figure}

Now we can readily see why the assignment $\R \mapsto {\cal A}_{\mathbb{Z}_2}(\R)$ violates additivity. 
Let $\R_1$ and $\R_2$ be two disjoint intervals, as in Figure \ref{fig:additivity}, and fix $\ell_1\in \R_1$ and $\ell_2\in \R_2$. 
Since $Z_{\ell_1}$ does not commute with $U$, it does not belong to ${\cal A}_{\mathbb{Z}_2}(\R_1)$. Similarly, we have $Z_{\ell_2}\notin {\cal A}_{\mathbb{Z}_2}(\R_2)$. 
However, $Z_{\ell_1}Z_{\ell_2}$ does commute with $U$, and belongs to ${\cal A}_{\mathbb{Z}_2}(\R_1\cup \R_2)$. Thus, ${\cal A}_{\mathbb{Z}_2}(\R_1\cup \R_2)\supsetneq {\cal A}_{\mathbb{Z}_2}(\R_1)\vee {\cal A}_{\mathbb{Z}_2}(\R_2),$ and additivity is violated by this pair of $\mathbb{Z}_2$-odd operators.

\begin{figure}
\centering
\begin{tikzpicture}[scale=0.7]  
    \draw[ultra thick] (0:2) arc[start angle=0, end angle=225, radius=2];

    \draw[ultra thick, red] (225:2) arc[start angle=225, end angle=360, radius=2];

    \foreach \angle in {0, 45, 90, 135, 180, 225, 270, 315} {
        \pgfmathsetmacro\x{2*cos(\angle)}
        \pgfmathsetmacro\y{2*sin(\angle)}
    \pgfmathsetmacro\dx{0.2*cos(\angle)}  
        \pgfmathsetmacro\dy{0.2*sin(\angle)}
        \draw[ultra thick] (\x-\dx, \y-\dy) -- (\x+\dx, \y+\dy);
    }

    \pgfmathsetmacro\x{1.6*cos(22.5)}
    \pgfmathsetmacro\y{1.6*sin(22.5)}
    \node at (\x, \y) {\footnotesize \textbf{$X_{\ell_1}$}};

    \pgfmathsetmacro\x{1.6*cos(-22.5)}
    \pgfmathsetmacro\y{1.6*sin(-22.5)}
    \node at (\x, \y) {\footnotesize \textbf{$X$}};
    
    \pgfmathsetmacro\x{1.6*cos(-67.5)}
    \pgfmathsetmacro\y{1.6*sin(-67.5)}
    \node at (\x, \y) {\footnotesize \textbf{$X$}};

    \pgfmathsetmacro\x{1.6*cos(-112.5)}
    \pgfmathsetmacro\y{1.6*sin(-112.5)}
    \node at (\x, \y) {\footnotesize \textbf{$X$}};
 
    \pgfmathsetmacro\x{1.6*cos(202.5)}
    \pgfmathsetmacro\y{1.6*sin(202.5)}
    \node at (\x, \y) {\footnotesize \textbf{$X_{\ell_2}$}};

    \node at (2, -1.7) {\footnotesize \textbf{\color{red}$\R$}};
    \node at (-1.7, 2) {\footnotesize \textbf{$\R'$}};

\end{tikzpicture}
\raisebox{1.5cm}{~~~~$\overset{\prod_\ell X_\ell=1}{=\joinrel=}$~~~~}
\begin{tikzpicture}[scale=0.7]  
    \draw[ultra thick] (0:2) arc[start angle=0, end angle=225, radius=2];

    \draw[ultra thick, red] (225:2) arc[start angle=225, end angle=360, radius=2];

    \foreach \angle in {0, 45, 90, 135, 180, 225, 270, 315} {
        \pgfmathsetmacro\x{2*cos(\angle)}
        \pgfmathsetmacro\y{2*sin(\angle)}
        
  \pgfmathsetmacro\dx{0.2*cos(\angle)}  
        \pgfmathsetmacro\dy{0.2*sin(\angle)}
        
        \draw[ultra thick] (\x-\dx, \y-\dy) -- (\x+\dx, \y+\dy);
    }

    \pgfmathsetmacro\x{1.6*cos(67.5)}
    \pgfmathsetmacro\y{1.6*sin(67.5)}
    \node at (\x, \y) {\footnotesize \textbf{$X$}};
    
    \pgfmathsetmacro\x{1.6*cos(112.5)}
    \pgfmathsetmacro\y{1.6*sin(112.5)}
    \node at (\x, \y) {\footnotesize \textbf{$X$}};

    \pgfmathsetmacro\x{1.6*cos(157.5)}
    \pgfmathsetmacro\y{1.6*sin(157.5)}
    \node at (\x, \y) {\footnotesize \textbf{$X$}};

    \node at (2, -1.7) {\footnotesize \textbf{\color{red}$\R$}};
    \node at (-1.7, 2) {\footnotesize \textbf{$\R'$}};

\end{tikzpicture}
\caption{Using the relation $U=\prod_{\ell=1}^L X_\ell=1$, we can rewrite the disorder operator  $U(\ell_1,\ell_2)$ in a presentation so that it is supported in $\R'$.}\label{fig:relation}
\end{figure}
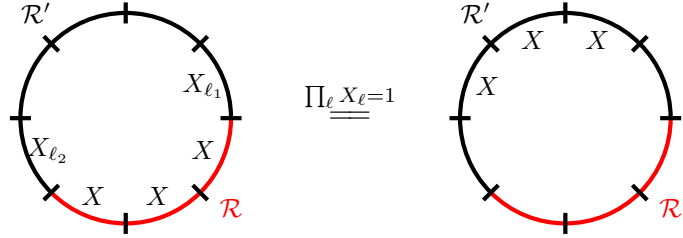

On the other hand, Haag duality is not violated for our choice of assignment. 
Naively, one might attempt to violate Haag duality by considering the following disorder operator \cite{Kadanoff:1970kz,Fradkin:2016ksx} from link $\ell_1$ to $\ell_2$:
\ie \label{disorder}
U(\ell_1,\ell_2)= \prod_{\ell= \ell_1 }^{\ell_2} X_\ell \,.
\fe
If we let $\R$ be the region from link $\ell_1+1$ to link $\ell_2-1$, then operator $U(\ell_1, \ell_2)$ commutes with every operator in ${\cal A}_{\mathbb{Z}_2}(\R)$, but it does not seem to be in ${\cal A}_{\mathbb{Z}_2}(\R')$ because of the appearance of $X_{\ell_1}$ and $X_{\ell_2}$.
This appears to be a violation of Haag duality, since it would imply $\A_{\mathbb{Z}_2}(\R)' \supsetneq \A_{\mathbb{Z}_2}(\R')$, but this is actually not the case. 
The reason is that on the $\mathbb{Z}_2$-even Hilbert space $\H_{\mathbb{Z}_2}$, we can rewrite \eqref{disorder} as
\ie
\label{eqn:X-Line}
U(\ell_1,\ell_2)= \prod_{\ell= \ell_2+1 }^{\ell_1+L-1} X_\ell\,.
\fe
Because $U(\ell_1, \ell_2)$ admits this presentation, we have $U(\ell_1,\ell_2)\in {\cal A}_{\mathbb{Z}_2}(\R')$, and Haag duality is not violated.
See Figure \ref{fig:relation}.

\begin{figure}
\centering
\begin{tikzpicture}[scale=0.7]  
    \draw[ultra thick] (45:2) arc[start angle=45, end angle=135, radius=2];

      \draw[ultra thick] (225:2) arc[start angle=225, end angle=315, radius=2];

    \draw[ultra thick, red] (315:2) arc[start angle=-45, end angle=45, radius=2];

    \draw[ultra thick, red] (135:2) arc[start angle=135, end angle=225, radius=2];

    \foreach \angle in {0, 45, 90, 135, 180, 225, 270, 315} {
        \pgfmathsetmacro\x{2*cos(\angle)}
        \pgfmathsetmacro\y{2*sin(\angle)}
        \pgfmathsetmacro\dx{0.2*cos(\angle)}
        \pgfmathsetmacro\dy{0.2*sin(\angle)}
        \draw[ultra thick] (\x-\dx, \y-\dy) -- (\x+\dx, \y+\dy);
    }

    \pgfmathsetmacro\x{1.6*cos(22.5)}
    \pgfmathsetmacro\y{1.6*sin(22.5)}
    \node at (\x, \y) {\footnotesize \textbf{$X_{\ell_1}$}};

    \pgfmathsetmacro\x{1.6*cos(-22.5)}
    \pgfmathsetmacro\y{1.6*sin(-22.5)}
    \node at (\x, \y) {\footnotesize \textbf{$X$}};
    
    \pgfmathsetmacro\x{1.6*cos(-67.5)}
    \pgfmathsetmacro\y{1.6*sin(-67.5)}
    \node at (\x, \y) {\footnotesize \textbf{$X$}};

    \pgfmathsetmacro\x{2.4*cos(-67.5)}
    \pgfmathsetmacro\y{2.4*sin(-67.5)}
    \node at (\x, \y) {\footnotesize \textbf{\color{blue}$Z$}};

    \pgfmathsetmacro\x{2.4*cos(112.5)}
    \pgfmathsetmacro\y{2.4*sin(112.5)}
    \node at (\x, \y) {\footnotesize \textbf{\color{blue}$Z$}};

    \pgfmathsetmacro\x{1.6*cos(-112.5)}
    \pgfmathsetmacro\y{1.6*sin(-112.5)}
    \node at (\x, \y) {\footnotesize \textbf{$X$}};
 
    \pgfmathsetmacro\x{1.6*cos(202.5)}
    \pgfmathsetmacro\y{1.6*sin(202.5)}
    \node at (\x, \y) {\footnotesize \textbf{$X_{\ell_2}$}};

    \node at (2.6, 0) {\footnotesize \textbf{\color{red}$\R$}};
    \node at (-2.6, 0) {\footnotesize \textbf{\color{red}$\R$}};
    \node at (0, 2.6) {\footnotesize \textbf{$\R'$}};
    \node at (0, -2.6) {\footnotesize \textbf{$\R'$}};
\end{tikzpicture}
\caption{The disorder operator $U(\ell_1,\ell_2)$ does not commute with a pair of $Z_\ell$'s in $\R'$.}\label{fig:Haag}
\end{figure}
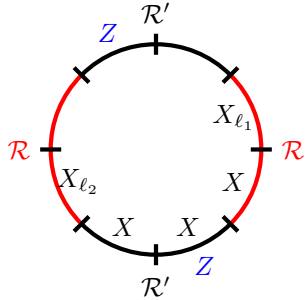

Another way one might attempt to violate Haag duality is by choosing $\R$ to be two disconnected intervals, with one component containing $\ell_1$ and the other containing $\ell_2$ (see Figure \ref{fig:Haag}).
Then $U(\ell_1,\ell_2)$ is not in $\A_{\mathbb{Z}_2}(\R)$, but it commutes with every operator of ${\cal A}_{\mathbb{Z}_2}(\R')$ that is supported entirely within a single component of $\R'$.
However, $U(\ell_1, \ell_2)$ fails to commute with a pair of $Z_\ell$ operators in the two disjoint intervals of $\R'$, so it is also not in ${\cal A}_{\mathbb{Z}_2}(\R')'$, and does not cause a violation of Haag duality.

By similar arguments as those given above, also discussed in \cite{Haag:book,Casini:2020rgj}, it is straightforward to see that similar conclusions hold in continuum QFT for any ordinary, invertible global symmetry $G$: the $G$-symmetric algebra assignment ${\cal A}_G$ violates additivity for disjoint regions because of the presence of charge-neutral pairs of charged local operators. 
In the next section we will see some qualitatively different features when we generalize this discussion to non-invertible symmetries.

\section{Symmetric sector under non-invertible symmetries}
\label{sec:continuum}

In this section we consider the symmetric sector of a CFT with respect to a  finite, internal, possibly non-invertible global symmetry.
We will see that, unlike the purely invertible case, additivity can be preserved in the symmetric sector in this more general setting.
The intuitive reason is that the ``charges'' of local operators under a non-invertible symmetry cannot generally be added.
Thus, one cannot generally violate additivity by creating a pair of charged local operators whose combined charge vanishes. 
In diagonal RCFTs, 
our general conclusion is that additivity is violated in the symmetric sector whenever the symmetry algebra contains a non-trivial invertible element, while Haag duality is violated whenever it contains a non-invertible element. 
(Here the symmetry of interest is the maximal internal global symmetry that commutes with the chiral algebra, generated by the Verlinde lines.) 
In addition to the example of section \ref{sec:invertible} where additivity was violated but Haag duality is preserved, in this section we find examples where additivity is preserved but Haag duality is violated, and examples where both additivity and Haag duality are violated.

\subsection{Kramers-Wannier symmetry in the Ising CFT}\label{sec:Ising}

\subsubsection{Review of Ising CFT and its symmetries}

The simplest example of non-invertible symmetry in QFT is the Kramers-Wannier symmetry in the $c=1/2$ Ising CFT \cite{Oshikawa:1996dj,Petkova:2000ip,Frohlich:2004ef,Chang:2018iay,Shao:2023gho}. 
The Ising CFT has three Virasoro local primary operators, which are the identity operator $1$, the energy operator $\epsilon$, and the order operator  $\sigma$. Their conformal weights are $(h,\bar h)=(0,0), (1/2,1/2),$ and $(1/16,1/16)$, respectively. Their fusion rule is
\ie
&\epsilon\times \epsilon=1\,,~~~\epsilon \times \sigma =\sigma \times \epsilon = \sigma\,,~~~\sigma\times \sigma = 1+\epsilon\,,
\fe
which means that the OPE channel between the two primary operators on the left-hand side contains the primaries on the right-hand side. 
The  lattice counterparts of $\sigma$ and $\epsilon$ are
\begin{equation}
    \sigma \sim Z_{\ell} \,,~~~
    \epsilon \sim Z_\ell Z_{\ell+1}  -X_{\ell+1}.
\end{equation}

The $\mathbb{Z}_2$ global symmetry of the Ising CFT is implemented by a topological line operator ${\cal L}_\epsilon$ that obeys ${\cal L}_\epsilon ^2=1$. 
(The notation will soon become clear.) 
This is the continuum counterpart of the operator $U=\prod_\ell X_\ell$ considered in Section \ref{sec:invertible}.
The commutators involving  ${\cal L}_\epsilon$ are ${\cal L}_\epsilon \, \epsilon   = \epsilon \, {\cal L}_\epsilon$ and ${\cal L}_\epsilon \,\sigma   = -\sigma \, {\cal L}_\sigma$. 

Interestingly, there is another global symmetry in the Ising CFT, implemented by a topological line operator ${\cal L}_\sigma$. 
Both ${\cal L}_\epsilon$ and ${\cal L}_\sigma$ commute with the stress-energy tensor, and therefore with the entire Virasoro algebra.
The  algebra of ${\cal L}_\epsilon$ and ${\cal L}_\sigma$ is the same as the fusion rule between the local primary operators:
\ie\label{Isingfusion}
&{\cal L}_\epsilon\times {\cal L}_\epsilon=1\,,~~~{\cal L}_\epsilon \times {\cal L}_\sigma ={\cal L}_\sigma \times {\cal L}_\epsilon = {\cal L}_\sigma\,,~~~{\cal L}_\sigma\times {\cal L}_\sigma = 1+{\cal L}_\epsilon\,.
\fe
More generally, in any diagonal rational conformal field theory, there is a one-to-one correspondence between the local primary operators and the topological lines, known as the Verlinde lines, that commute with the extended chiral algebra and that obey the same fusion rule as the primaries \cite{Verlinde:1988sn,Petkova:2000ip,Drukker:2010jp,Chang:2018iay}.
This more general setting is discussed in Section \ref{sec:RCFT}.

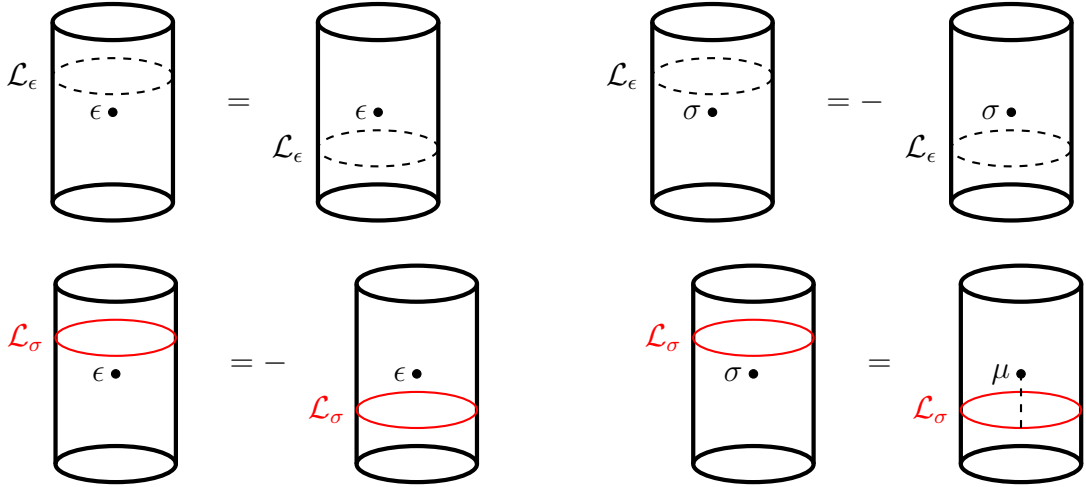
\begin{figure}
\centering
\begin{tikzpicture}[scale=.8]
    \draw[ultra thick] (0,1.5) ellipse (1 and 0.3);
    
    \draw[ultra thick] (-1,1.5) -- (-1,-1.5);
    \draw[ultra thick] (1,1.5) -- (1,-1.5);
    
    \draw[ultra thick] (0,-1.5) ellipse (1 and 0.3);
    
    \draw[black, thick, dashed] (0,0.6) ellipse (1 and 0.3);

        \node at (-1.5,0.6) {\textbf{${\cal L}_\epsilon$}};
        
\filldraw[black] (0,0) circle (2pt) node[anchor=east]{$\epsilon$};

\end{tikzpicture}
\raisebox{1.5cm}{~~~~$=$}
\begin{tikzpicture}[scale=.8]
    \draw[ultra thick] (0,1.5) ellipse (1 and 0.3);
    
    \draw[ultra thick] (-1,1.5) -- (-1,-1.5);
    \draw[ultra thick] (1,1.5) -- (1,-1.5);
    
    \draw[ultra thick] (0,-1.5) ellipse (1 and 0.3);
    
    \draw[black, thick, dashed] (0,-.6) ellipse (1 and 0.3);

        \node at (-1.5,-.6) {\textbf{${\cal L}_\epsilon$}};
        
\filldraw[black] (0,0) circle (2pt) node[anchor=east]{$\epsilon$};
\end{tikzpicture}
~~~~~~~~~~~~~
\begin{tikzpicture}[scale=.8]
    \draw[ultra thick] (0,1.5) ellipse (1 and 0.3);
    
    \draw[ultra thick] (-1,1.5) -- (-1,-1.5);
    \draw[ultra thick] (1,1.5) -- (1,-1.5);
    
    \draw[ultra thick] (0,-1.5) ellipse (1 and 0.3);
    
    \draw[black, thick, dashed] (0,0.6) ellipse (1 and 0.3);

        \node at (-1.5,0.6) {\textbf{${\cal L}_\epsilon$}};
        
\filldraw[black] (0,0) circle (2pt) node[anchor=east]{$\sigma$};

\end{tikzpicture}
\raisebox{1.5cm}{~~~~$= -$}
\begin{tikzpicture}[scale=.8]
    \draw[ultra thick] (0,1.5) ellipse (1 and 0.3);
    
    \draw[ultra thick] (-1,1.5) -- (-1,-1.5);
    \draw[ultra thick] (1,1.5) -- (1,-1.5);
    
    \draw[ultra thick] (0,-1.5) ellipse (1 and 0.3);
    
    \draw[black, thick, dashed] (0,-.6) ellipse (1 and 0.3);

        \node at (-1.5,-.6) {\textbf{${\cal L}_\epsilon$}};
        
\filldraw[black] (0,0) circle (2pt) node[anchor=east]{$\sigma$};
\end{tikzpicture}
~\\
~\\
\begin{tikzpicture}[scale=.8]
    \draw[ultra thick] (0,1.5) ellipse (1 and 0.3);
    
    \draw[ultra thick] (-1,1.5) -- (-1,-1.5);
    \draw[ultra thick] (1,1.5) -- (1,-1.5);
    
    \draw[ultra thick] (0,-1.5) ellipse (1 and 0.3);
    
    \draw[red, thick] (0,0.6) ellipse (1 and 0.3);

        \node at (-1.5,0.6) {\color{red}\textbf{${\cal L}_\sigma$}};
        
\filldraw[black] (0,0) circle (2pt) node[anchor=east]{$\epsilon$};

\end{tikzpicture}
\raisebox{1.5cm}{~~~~$= -$}
\begin{tikzpicture}[scale=.8]
    \draw[ultra thick] (0,1.5) ellipse (1 and 0.3);
    
    \draw[ultra thick] (-1,1.5) -- (-1,-1.5);
    \draw[ultra thick] (1,1.5) -- (1,-1.5);
    
    \draw[ultra thick] (0,-1.5) ellipse (1 and 0.3);
    
    \draw[red, thick] (0,-.6) ellipse (1 and 0.3);

        \node at (-1.5,-.6) {\color{red}\textbf{${\cal L}_\sigma$}};
        
\filldraw[black] (0,0) circle (2pt) node[anchor=east]{$\epsilon$};
\end{tikzpicture}
~~~~~~~~~~~~~
\begin{tikzpicture}[scale=.8]
    \draw[ultra thick] (0,1.5) ellipse (1 and 0.3);
    
    \draw[ultra thick] (-1,1.5) -- (-1,-1.5);
    \draw[ultra thick] (1,1.5) -- (1,-1.5);
    
    \draw[ultra thick] (0,-1.5) ellipse (1 and 0.3);
    
    \draw[red, thick] (0,0.6) ellipse (1 and 0.3);

        \node at (-1.5,0.6) {\color{red}\textbf{${\cal L}_\sigma$}};
        
\filldraw[black] (0,0) circle (2pt) node[anchor=east]{$\sigma$};

\end{tikzpicture}
\raisebox{1.5cm}{~~~~$=$}
\begin{tikzpicture}[scale=.8]
    \draw[ultra thick] (0,1.5) ellipse (1 and 0.3);
    
    \draw[ultra thick] (-1,1.5) -- (-1,-1.5);
    \draw[ultra thick] (1,1.5) -- (1,-1.5);
    
    \draw[ultra thick] (0,-1.5) ellipse (1 and 0.3);
    
    \draw[red, thick] (0,-.6) ellipse (1 and 0.3);

\node at (-1.5,-.6) {\textbf{\color{red}${\cal L}_\sigma$}};

\draw[thick,dashed] (0,-.9) -- (0,0);

\filldraw[black] (0,0) circle (2pt) node[anchor=east]{$\mu$};
\end{tikzpicture}
\caption{The commutation relation between the invertible $\mathbb{Z}_2$ topological line operator ${\cal L}_\epsilon$ (dashed line), the non-invertible Kramers-Wannier line operator ${\cal L}_\sigma$ (red line), and the local primaries $\epsilon$ and $\sigma$ \cite{Frohlich:2004ef}. }\label{fig:cylinder}
\end{figure}

The actions of ${\cal L}_\epsilon$ and ${\cal L}_\sigma$ on the Virasoro primary states on a circle are given by
\ie
&{\cal L}_\epsilon \ket{1} = \ket{1} \,,~~~&&{\cal L}_\epsilon \ket{\epsilon} = \ket{\epsilon}\,,~~~&&{\cal L}_\epsilon \ket{\sigma} =- \ket{\sigma}\,,\\
&{\cal L}_\sigma \ket{1} = \sqrt{2}\ket{1} \,,~~~&&{\cal L}_\sigma \ket{\epsilon} = -\sqrt{2}\ket{\epsilon}\,,~~~&&{\cal L}_\sigma \ket{\sigma} =0\,.
\fe
Here $\ket{1},\ket{\epsilon},$ and $\ket{\sigma}$ are the Virasoro primary states in the (untwisted) Hilbert space that correspond to the local primary operators $1, \epsilon,$ and $\sigma$ via the operator-state correspondence. 
The operator ${\cal L}_\sigma$ has a kernel and is therefore non-invertible.
It generates a non-invertible  global symmetry, which is associated with the Kramers-Wannier transformation \cite{PhysRev.60.252}.
The algebraic structure of these topological lines is not that of a group, but of a fusion category --- this structure includes both the symmetry algebra in equation \eqref{Isingfusion} and additional information about junctions where distinct lines meet.
The category we are studying here is usually called the ``Ising fusion category.''
We will not list all of its categorical data explicitly, but some information beyond the symmetry algebra is needed to describe the commutation relations between topological lines and local primaries, which can be found in \cite{Frohlich:2004ef,Bhardwaj:2017xup,Chang:2018iay} and are shown in Figure \ref{fig:cylinder}.

While ${\cal L}_\sigma$ anticommutes with $\epsilon,$ it does not have a simple commutation relation with $\sigma.$
Instead, as shown in Figure \ref{fig:cylinder}, commuting $\L_{\sigma}$ past $\sigma$ produces a ``disorder operator" $\mu$ whose conformal weight is $(h,\bar h)=(\frac{1}{16},\frac{1}{16})$.
The object $\mu$ is a ``twisted'' primary operator that lives at the end of the line ${\cal L}_\epsilon$, and in Figure \ref{fig:cylinder}, this line $\L_{\epsilon}$ attaches to $\L_{\sigma}$ via a junction.

More generally, there are three ``twisted'' primary operators that can live at the end of  ${\cal L}_\epsilon$, denoted $\mu,\psi,$ and $\bar \psi$, with conformal weights $(h,\bar h)=(\frac{1}{16},\frac{1}{16}), (\frac12,0 ),$ and $(0,\frac12)$. 
The twisted primaries $\psi(z)$ and $\bar\psi(\bar z)$ are left- and right-moving free fermion fields, which are not local operators in the Ising CFT.
These operators are said to be ``twisted'' because via the state-operator map, they correspond to states in a Hilbert space on the cylinder that has been twisted by an ${\cal L}_\epsilon$ defect.

It is a common abuse of terminology to refer to $\mu$ as the disorder ``operator", even though it is not an operator  acting on the  Hilbert space. On the other hand, a pair of $\mu$'s connected by a $\mathbb{Z}_2$ line ${\cal L}_\epsilon$ does act on the Hilbert space.
Such a pair is the continuum counterpart of the lattice disorder operator $U(\ell_1,\ell_2)=\prod_{\ell=\ell_1}^{\ell_2}X_\ell$, which we introduced previously in equation \eqref{disorder}.  
The OPE between any pair of the twisted primaries $\mu,\psi,$ or $\bar\psi$ gives a sum of local operators, which we record below \cite{Ginsparg:1988ui}:
\ie\label{twistedOPE}
&\psi\times \psi =\bar\psi\times \bar\psi =1\,,~~~&&\psi\times\bar \psi =\bar\psi \times \psi =\epsilon\,,\\
&\psi \times \mu = \mu\times \psi  = \sigma\,,~~~
&&\bar\psi \times \mu = \mu\times \bar\psi  = \sigma\,,~~~
&&\mu \times \mu = 1+\epsilon\,.
\fe

\subsubsection{Symmetric sectors}
\label{subsec:symmetric-sectors}

Now we can consider different sectors of the Ising CFT relative to invertible or non-invertible symmetries. 
The $\mathbb{Z}_2$-even sector ${\cal A}_{\mathbb{Z}_2}$ of the Ising CFT consists of the local primaries $1$ and $\epsilon$, as well as their descendants. 
As explained in Section \ref{sec:invertible}, we find that a pair of the order operators $\sigma$  violates additivity in  ${\cal A}_{\mathbb{Z}_2}$, whereas Haag duality is respected.

Next, we consider the symmetric sector ${\cal A}_\text{KW}$ under the full Ising fusion category. 
More specifically, ${\cal A}_\text{KW}$ is defined as the set of operators that commute with both ${\cal L}_\epsilon$ and ${\cal L}_\sigma$, i.e., any ${\cal O}$ that  satisfies ${\cal O} {\cal L}_\epsilon = {\cal L}_\epsilon {\cal O} ,~{\cal O} {\cal L}_\sigma = {\cal L}_\sigma {\cal O}$. 
As far as local operators are concerned, this sector consists only of the identity primary $1$ and its descendants such as the stress-energy tensor. 
Below we show that this sector ${\cal A}_\text{KW}$ violates both additivity and Haag duality.

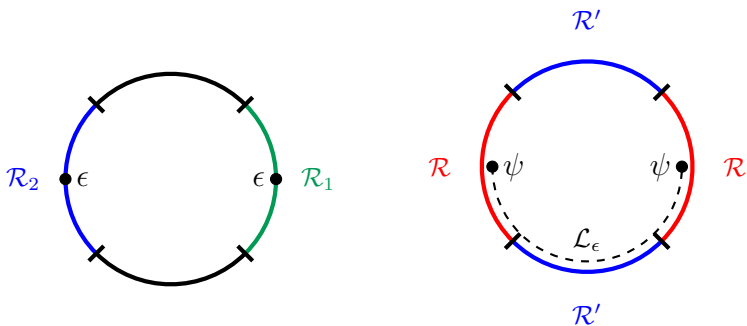
\begin{figure}[h!]
\centering
\raisebox{.5cm}{\begin{tikzpicture}[scale=0.7]  
    \draw[ultra thick] (45:2) arc[start angle=45, end angle=135, radius=2];

      \draw[ultra thick] (225:2) arc[start angle=225, end angle=315, radius=2];

    \draw[ultra thick, ForestGreen] (315:2) arc[start angle=-45, end angle=45, radius=2];

    \draw[ultra thick, blue] (135:2) arc[start angle=135, end angle=225, radius=2];

    \foreach \angle in { 45, 135,  225, 315} {
        \pgfmathsetmacro\x{2*cos(\angle)}
        \pgfmathsetmacro\y{2*sin(\angle)}
        
        \pgfmathsetmacro\dx{0.2*cos(\angle)}  
        \pgfmathsetmacro\dy{0.2*sin(\angle)}
        
        \draw[ultra thick] (\x-\dx, \y-\dy) -- (\x+\dx, \y+\dy);
    }

\filldraw[black] (-2,0) circle (3pt) node[anchor=west]{$\epsilon$};

\filldraw[black] (2,0) circle (3pt) node[anchor=east]{$\epsilon$};

    \node at (2.8, 0) {\footnotesize \textbf{\color{ForestGreen}$\R_1$}};
    \node at (-2.8, 0) {\footnotesize \textbf{\color{blue}$\R_2$}};

\end{tikzpicture}}
~~~~~
\begin{tikzpicture}[scale=0.7]  
    \draw[ultra thick,blue] (45:2) arc[start angle=45, end angle=135, radius=2];

      \draw[ultra thick,blue] (225:2) arc[start angle=225, end angle=315, radius=2];

    \draw[ultra thick, red] (315:2) arc[start angle=-45, end angle=45, radius=2];

    \draw[ultra thick, red] (135:2) arc[start angle=135, end angle=225, radius=2];

     \draw[ thick, black,dashed] (180:1.8) arc[start angle=180, end angle=360, radius=1.8];

    \foreach \angle in { 45,  135,  225,  315} {
        \pgfmathsetmacro\x{2*cos(\angle)}
        \pgfmathsetmacro\y{2*sin(\angle)}
        
        \pgfmathsetmacro\dx{0.2*cos(\angle)}  
        \pgfmathsetmacro\dy{0.2*sin(\angle)}
        
        \draw[ultra thick] (\x-\dx, \y-\dy) -- (\x+\dx, \y+\dy);
    }

  \filldraw[black] (-1.8,0) circle (3pt) node[anchor=west]{$\psi$};

\filldraw[black] (1.8,0) circle (3pt) node[anchor=east]{$\psi$};

   \node at (0, -1.4) {\footnotesize \textbf{${\cal L}_\epsilon$}};

    \node at (2.8, 0) {\footnotesize \textbf{\color{red}$\R$}};
    \node at (-2.8, 0) {\footnotesize \textbf{\color{red}$\R$}};
    \node at (0, 2.8) {\footnotesize \textbf{\color{blue}$\R'$}};
    \node at (0, -2.8) {\footnotesize \textbf{\color{blue}$\R'$}};
\end{tikzpicture}
\caption{The symmetric sector with respect to the full Ising fusion category in the Ising CFT violates both additivity and Haag duality. Left: A pair of the energy operators $\epsilon$ (with $(h,\bar h)=(\frac12,\frac12)$) violates additivity. Right: 
A pair of left-moving free fermions $\psi$ (with $(h,\bar h)=(\frac12,0)$) connected by a $\mathbb{Z}_2$ line ${\cal L}_\epsilon$ (shown in dashed line) violates Haag duality. }\label{fig:Ising}
\end{figure}

To violate additivity, we note that even though a single $\epsilon$ primary operator does not belong to ${\cal A}_\text{KW}$, a pair of them commutes with both ${\cal L}_\epsilon$ and ${\cal L}_\sigma$. 
If we place one $\epsilon$ in an interval $\R_1$ and another $\epsilon$ in a distinct interval $\R_2$, as in Figure \ref{fig:Ising}, we find that this bilocal operator belongs to ${\cal A}_\text{KW}(\R_1\cup \R_2)$ but is not in ${\cal A}_\text{KW}(\R_1)\vee{\cal A}_\text{KW}(\R_2)$. Thus, additivity is violated. 

Haag duality is violated by considering a pair of left-moving free fermion operators $\psi(z)$ (whose  conformal weights are $(h,\bar h)=(1/2,0)$) connected by ${\cal L}_\epsilon$. 
The OPE between two free fermions is $\psi(z) \psi(0)\sim 1/z$, which gives us the fusion rule $\psi \times\psi =1$ from equation \eqref{twistedOPE}.
Based on this OPE, we therefore argue that a pair of $\psi$ belongs to ${\cal A}_\text{KW}$. (See, however,  more discussions below.) 
Now we place one $\psi$ in one disconnected component of the region $\R$ in  Figure \ref{fig:Ising}, and another $\psi$ in the other component, then the resulting bilocal operator is not contained in  ${\cal A}_\text{KW}(\R)$ because of the ${\cal L}_\epsilon$ line connecting the fermions. 
However, the $\psi\psi$ bilocal operator commutes with every operator in $\R'$, and is therefore  in ${\cal A}_\text{KW}(\R')'$.
To see this, we note that every operator in $\R'$ is a multilocal combination of primary or descendant operators that commutes with both $\L_{\sigma}$ and $\L_{\epsilon}.$
But no multilocal operator containing $\sigma$ can be in $\A_{\text{KW}}(\R')$ --- not even the $\sigma \sigma$ pair that we used to rescue Haag duality in section \ref{sec:invertible} --- because commuting $\L_{\sigma}$ past $\sigma$ introduces a disorder operator $\mu$ that cannot be cancelled by the addition of other local terms. 
So all the operators in $\A_{\text{KW}}$ are generated  the identity,  $\epsilon \epsilon$ pairs, and their descendants,  but these commute with a pair of $\psi$.
We therefore see that the symmetric sector of the Ising fusion category violates Haag duality in the form $\A_{\text{KW}}(\R')' \supsetneq \A_{\text{KW}}(\R).$

\begin{figure}
\centering
\begin{tikzpicture}[scale=.8] 
    \draw[ultra thick] (0,1.5) ellipse (1.5 and 0.4); 
    
    \draw[ultra thick] (-1.5,1.5) -- (-1.5,-1.5);
    \draw[ultra thick] (1.5,1.5) -- (1.5,-1.5);
    
    \draw[ultra thick] (0,-1.5) ellipse (1.5 and 0.4); 
    
    \draw[red, thick] (0,0.6) ellipse (1.5 and 0.4); 

    \node at (-2,0.6) {\color{red}\textbf{${\cal L}_\sigma$}};
        
    \filldraw[black] (-.7,-.2) circle (2pt) node[anchor=east]{$\sigma$};
    \filldraw[black] (0.7,-0.2) circle (2pt) node[anchor=west]{$\sigma$};
\end{tikzpicture}
\raisebox{1.5cm}{~~~~$=$}
\begin{tikzpicture}[scale=.8] 
    \draw[ultra thick] (0,1.5) ellipse (1.5 and 0.4); 
    
    \draw[ultra thick] (-1.5,1.5) -- (-1.5,-1.5);
    \draw[ultra thick] (1.5,1.5) -- (1.5,-1.5);
    
    \draw[ultra thick] (0,-1.5) ellipse (1.5 and 0.4);

    \draw[red, thick] (0,-.8) ellipse (1.5 and 0.4);

    \node at (-2,-.8) {\textbf{\color{red}${\cal L}_\sigma$}};

    \draw[thick,dashed] (-.7,-.1) -- (-0.7,-1.15) ;
     \draw[thick,dashed] (.7,-.1) -- (0.7,-1.15) ;

\filldraw[black] (-.7,-.1) circle (2pt) node[anchor=east]{$\mu$};
\filldraw[black] (0.7,-0.1) circle (2pt) node[anchor=west]{$\mu$};

\end{tikzpicture}
\raisebox{1.5cm}{~~~~$=$}
\begin{tikzpicture}[scale=.8] 
    \draw[ultra thick] (0,1.5) ellipse (1.5 and 0.4);
    
    \draw[ultra thick] (-1.5,1.5) -- (-1.5,-1.5);
    \draw[ultra thick] (1.5,1.5) -- (1.5,-1.5);
    
    \draw[ultra thick] (0,-1.5) ellipse (1.5 and 0.4); 
    
    \draw[red, thick] (0,-.8) ellipse (1.5 and 0.4); 

    \node at (-2,-.8) {\textbf{\color{red}${\cal L}_\sigma$}};

    \draw[thick,dashed] (-.7,-.1) -- (0.7,-0.1) ;
    \node at (0,0.3) {\footnotesize\textbf{${\cal L}_\epsilon$}};

\filldraw[black] (-.7,-.1) circle (2pt) node[anchor=east]{$\mu$};
\filldraw[black] (0.7,-0.1) circle (2pt) node[anchor=west]{$\mu$};

\end{tikzpicture}

\caption{The commutation relation between the  non-invertible  operator ${\cal L}_\sigma$ (red line), and a pair of order operators  $\sigma$. The non-invertible operator ${\cal L}_\sigma$ implements a Kramers-Wannier transformation and turns a pair of order operators into 
  a pair of disorder operators connected by a $\mathbb{Z}_2$ line ${\cal L}_\epsilon$ \cite{Frohlich:2004ef}. Therefore, a pair of $\sigma$'s belongs to ${\cal A}_{\mathbb{Z}_2}$ but not ${\cal A}_\text{KW}$. (See \eqref{latticeKW} for the lattice counterpart of this equation.)}\label{fig:KW}
\end{figure}
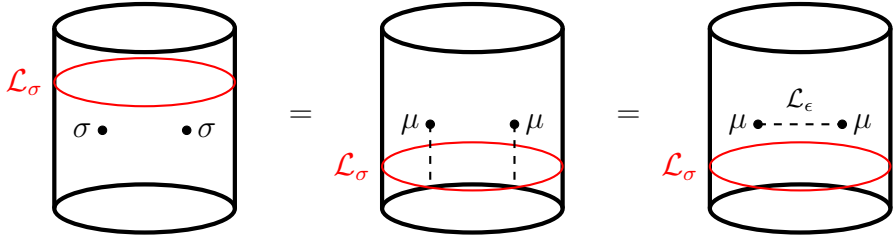

In the above argument, we used the OPE to show that $\psi(z)\psi(0)$ belongs to the symmetric sector of the Ising fusion category. 
One might worry that this reasoning  is only valid when $z$ is small and when   no other operator insertions are present. 
However,  we claim that this argument extends even to finite $z$. 
The key is that the charges (or representations) of a finite, possibly non-invertible,  global symmetry are discrete, and cannot be continuously deformed. 
For instance, a $\mathbb{Z}_2$-odd charged operator cannot suddenly become $\mathbb{Z}_2$-even under a continuous deformation. 
In the Ising CFT, we  first establish that $\psi(z)\psi(0)$ commutes with the Ising fusion category when $z$ is small using the OPE, then use the rigidity of the discrete representations to conclude that $\psi(z) \psi(0)$ remains in ${\cal A}_\text{KW}$ even at finite separation $z$.

As a tangent, we note that there is an alternative way to see why a pair of left-moving fermions $\psi$'s is symmetric under the full Ising fusion category, without relying on the OPE. 
The Ising CFT is closely related to the free Majorana CFT, however, they are globally two different QFTs.
The Majorana CFT has an invertible $\mathbb{Z}_2\times \mathbb{Z}_2$ global symmetry, generated by a chiral fermion parity $(-1)^{F_\text{L}}$ that flips the sign of the left-moving fermion $\psi(z)$, and a non-chiral fermion parity $(-1)^F$ that flips the sign of both the left- and right-moving fermions $\psi(z),\bar \psi(\bar z)$. The Ising CFT is obtained by gauging $(-1)^F$ of the Majorana CFT (see, for example,  \cite{Karch:2019lnn} for a review).
Gauging a finite symmetry produces a topological Wilson line in the gauged theory, and in the case of $(-1)^F$ the corresponding Wilson line is the $\mathbb{Z}_2$ line $\mathcal{L}_{\epsilon}$ of the Ising CFT.
Since the left- and right-moving free fermions $\psi(z)$ and $\bar \psi(\bar  z)$ were charged under $(-1)^F$, they are now connected to the Wilson line ${\cal L}_\epsilon$ in the Ising CFT. 
More interestingly, because of an 't Hooft anomaly between $(-1)^{F_\text{L}}$ and $(-1)^F$  \cite{Delmastro:2021xox}, the chiral fermion parity turns into the non-invertible global symmetry ${\cal L}_\sigma$ in the Ising CFT \cite{Ji:2019ugf,Lin:2019hks,Kaidi:2021xfk,Seiberg:2023cdc}. 
Since ${\cal L}_\sigma$ originates from $(-1)^{F_\text{L}}$, it commutes with a pair of the left-moving fermions $\psi$. 
In contrast, a left-moving fermion $\psi $ connected to a right-moving fermion $\bar\psi $ by a ${\cal L}_\epsilon$ line is odd under ${\cal L}_\sigma$, consistent with the OPE $\psi \times \bar \psi =\epsilon$.

\subsection{Fibonacci symmetry}\label{sec:Fib}

In this subsection we consider RCFTs with only one nontrivial local primary operator with respect to the extended chiral algebra. 
These include the $c=14/5$ $(\mathfrak{g}_2)_1$ and  the $c=26/5$ $(\mathfrak{f}_4)_1$ WZW  models.
The final conclusion also holds for the non-unitary $c=-22/5$ Lee-Yang CFT (which is the (2,5) minimal model), but some of the details differ. 
(See, for example,  \cite{DiFrancesco:1997nk} for reviews of minimal models and WZW models $(\mathfrak{h})_k$ at level $k$ based on a Lie algebra $\mathfrak{h}$.)

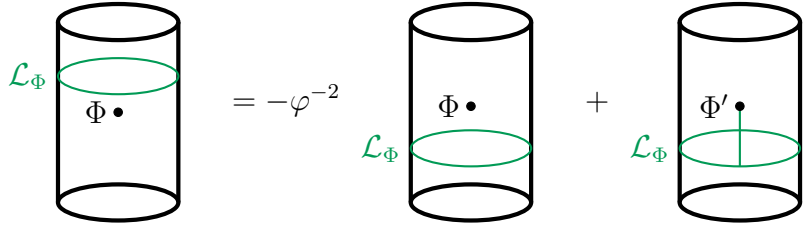
\begin{figure}
\centering
\begin{tikzpicture}[scale=.8]
    \draw[ultra thick] (0,1.5) ellipse (1 and 0.3);
    
    \draw[ultra thick] (-1,1.5) -- (-1,-1.5);
    \draw[ultra thick] (1,1.5) -- (1,-1.5);
    
    \draw[ultra thick] (0,-1.5) ellipse (1 and 0.3);
    
    \draw[ForestGreen, thick] (0,0.6) ellipse (1 and 0.3);

        \node at (-1.5,0.6) {\color{ForestGreen}\textbf{${\cal L}_\Phi$}};
        
\filldraw[black] (0,0) circle (2pt) node[anchor=east]{$\Phi$};

\end{tikzpicture}
\raisebox{1.5cm}{~~~~$=-\varphi^{-2}$}
\begin{tikzpicture}[scale=.8]
    \draw[ultra thick] (0,1.5) ellipse (1 and 0.3);
    
    \draw[ultra thick] (-1,1.5) -- (-1,-1.5);
    \draw[ultra thick] (1,1.5) -- (1,-1.5);
    
    \draw[ultra thick] (0,-1.5) ellipse (1 and 0.3);
    
    \draw[ForestGreen, thick] (0,-.6) ellipse (1 and 0.3);

\node at (-1.5,-.6) {\textbf{\color{ForestGreen}${\cal L}_\Phi$}};

\filldraw[black] (0,0.1) circle (2pt) node[anchor=east]{$\Phi$};
\end{tikzpicture}
\raisebox{1.5cm}{~~~~$+$}
\begin{tikzpicture}[scale=.8]
    \draw[ultra thick] (0,1.5) ellipse (1 and 0.3);
    
    \draw[ultra thick] (-1,1.5) -- (-1,-1.5);
    \draw[ultra thick] (1,1.5) -- (1,-1.5);
    
    \draw[ultra thick] (0,-1.5) ellipse (1 and 0.3);
    
    \draw[ForestGreen, thick] (0,-.6) ellipse (1 and 0.3);

\node at (-1.5,-.6) {\textbf{\color{ForestGreen}${\cal L}_\Phi$}};

\draw[thick,ForestGreen] (0,-.9) -- (0,0.1);

\filldraw[black] (0,0.1) circle (2pt) node[anchor=east]{$\Phi'$};
\end{tikzpicture}
\caption{The commutation relation between the Fibonacci topological line operator ${\cal L}_\Phi$ (green line) and the local primary operator $\Phi$. Here $\Phi'$ is a twisted primary with conformal weights $(h,\bar h)=(\frac25,\frac25)$ and $\varphi=\frac{1+\sqrt{5}}{2}$ is the golden ratio. (In the second term we have normalized the junction and $\Phi'$ so that the coefficient is $+1$.)}\label{fig:Fib}
\end{figure}

For concreteness, we focus on the $(\mathfrak{g}_2)_1$ WZW model.
It has two current algebra primaries, the identity operator 1 and a nontrivial primary $\Phi$. 
Their conformal weights are $(h,\bar h)=(0,0)$ and $(h,\bar h)=(\frac25,\frac25)$, respectively. The fusion  rule is  \cite{Verlinde:1988sn}
\ie
\Phi\times \Phi=1+\Phi\,.
\fe
The CFT has a non-invertible topological line ${\cal L}_\Phi$ that commutes with the $(\mathfrak{g}_2)_1$ Kac-Moody current algebra \cite{2007arXiv0712.1377R,Lin:2023uvm}.
This line operator obeys the same fusion rule:
\ie
{\cal L}_\Phi \times {\cal L}_\Phi = 1+{\cal L}_\Phi\,,
\fe
and forms the Fibonacci fusion category \cite{Moore:1988qv,Freedman:2006yr,2009arXiv0902.3275T,Chang:2018iay}, which takes its name from the similarity between the fusion rule and the golden ratio formula $\varphi^2 = 1 + \varphi$. 
The operator $\L_{\Phi}$ is called the Fibonacci symmetry; it commutes with 1, but does not commute with $\Phi$.  The commutator between ${\cal L}_\Phi$ and $\Phi$ is shown in Figure \ref{fig:Fib}, and can be derived using detailed properties of the fusion category as in \cite{Chang:2018iay}.

There are three twisted primaries that can live at the end of a Fibonacci line  ${\cal L}_\Phi$, which we denote by $\Phi', \Psi,$ and $\bar\Psi$.
Their conformal weights are $(h,\bar h)=\left(\frac25,\frac25\right) , \left( \frac25,0\right),$ and $\left( 0,\frac25\right)$.
The OPEs between these operators are
\ie
&\Psi \times \Psi = \bar\Psi \times \bar\Psi =1\,,~~~&& \Psi \times \bar\Psi =\bar\Psi \times \Psi = \Phi\,,\\
&\Psi \times \Phi' = \Phi'\times \Psi = \Phi\,,~~~&&\bar\Psi \times \Phi' = \Phi'\times \bar\Psi = \Phi\,,~~~&&
\Phi'\times \Phi' = 1+ \Phi\,.
\fe
From the OPEs,  and from the argument given at the end of Subsection \ref{subsec:symmetric-sectors}, we see that the twisted primaries $\Psi \Psi$ and $\bar{\Psi} \bar{\Psi}$  both commute with ${\cal L}_\Phi$.

The Fibonacci-symmetric subalgebra  ${\cal A}_\text{Fib}$ consists only of the identity sector, which is generated by the Kac-Moody currents and their descendants.
For a violation of additivity to occur, we would need to find a bilocal operator that commutes with the Fibonacci category.
But no such operator exists, since commuting $\L_{\Phi}$ past $\Phi$ introduces a nontrivial twisted primary $\Phi'$.
From Figure \ref{fig:Fib}, we see that no combination of $\Phi$ operators or descendants can commute with $\L_{\Phi}$.
Put differently, the ``charges'' of the only nontrivial primary $\Phi$ are not complex numbers, and cannot be made to add up to zero by putting multiple $\Phi$ operators together.
Since no such bilocal operators exist, the algebra $\A_{\text{Fib}}(\R)$ for any region $\R$ consists only of Kac-Moody currents and descendants contained in $\R$.
We conclude that ${\cal A}_\text{Fib}$ respects additivity.

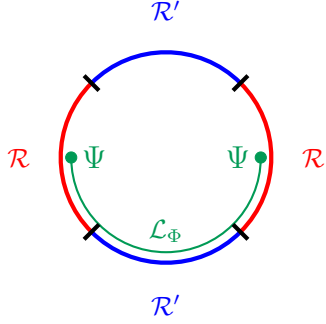
\begin{figure}
\centering
\begin{tikzpicture}[scale=0.7]  
    \draw[ultra thick,blue] (45:2) arc[start angle=45, end angle=135, radius=2];

      \draw[ultra thick,blue] (225:2) arc[start angle=225, end angle=315, radius=2];

    \draw[ultra thick, red] (315:2) arc[start angle=-45, end angle=45, radius=2];

    \draw[ultra thick, red] (135:2) arc[start angle=135, end angle=225, radius=2];

     \draw[ thick, ForestGreen] (180:1.8) arc[start angle=180, end angle=360, radius=1.8];

    \foreach \angle in { 45,  135,  225,  315} {
        \pgfmathsetmacro\x{2*cos(\angle)}
        \pgfmathsetmacro\y{2*sin(\angle)}
        
        \pgfmathsetmacro\dx{0.2*cos(\angle)}  
        \pgfmathsetmacro\dy{0.2*sin(\angle)}
        
        \draw[ultra thick] (\x-\dx, \y-\dy) -- (\x+\dx, \y+\dy);
    }

  \filldraw[ForestGreen] (-1.8,0) circle (3pt) node[anchor=west]{\color{ForestGreen}$\Psi$};

\filldraw[ForestGreen] (1.8,0) circle (3pt) node[anchor=east]{\color{ForestGreen}$\Psi$};

   \node at (0, -1.4) {\footnotesize \textbf{\color{ForestGreen}${\cal L}_\Phi$}};

    \node at (2.8, 0) {\footnotesize \textbf{\color{red}$\R$}};
    \node at (-2.8, 0) {\footnotesize \textbf{\color{red}$\R$}};
    \node at (0, 2.8) {\footnotesize \textbf{\color{blue}$\R'$}};
    \node at (0, -2.8) {\footnotesize \textbf{\color{blue}$\R'$}};
\end{tikzpicture}
\caption{A pair of  $\Psi$ (with $(h,\bar h)=(\frac25,0)$) connected by  ${\cal L}_\Phi$  violates Haag duality in the symmetric sector of  the $(\mathfrak{g}_2)_1$ WZW model with respect to the Fibonacci fusion category. 
This operator is in ${\cal A}_\text{Fib}({\cal R}')'$ but not in ${\cal A}_\text{Fib}({\cal R})$. }\label{fig:FibHaag}
\end{figure}

However, we can easily show that $\A_{\text{Fib}}$ violates Haag duality.
To see this, consider $\Psi \Psi$, which consists of a pair of twisted $\Psi$ primaries connected by a Fibonacci line ${\cal L}_\Phi$ as in Figure \ref{fig:FibHaag}.
If $\R$ contains two disjoint intervals and the endpoints of $\Psi \Psi$ are placed in different intervals, then $\Psi \Psi$ is not contained in the algebra $\A_{\text{Fib}}(\R)$.
However, because we have just argued that $\A_{\text{Fib}}(\R')$ only contains descendants of Kac-Moody currents, we see that $\Psi \Psi$ commutes with every operator in $\A_{\text{Fib}}(\R').$
This violates Haag duality in the form $\A_{\text{Fib}}(\R')' \supsetneq \A_{\text{Fib}}(\R).$

\subsection{Verlinde lines in a diagonal RCFT}\label{sec:RCFT}

We have seen in Sections \ref{sec:Ising} and \ref{sec:Fib} that the symmetric sectors of two different non-invertible symmetry algebras violate additivity and Haag duality in different ways. 
Here we extend these discussions to general (bosonic) diagonal RCFTs.
See \cite{Ginsparg:1988ui,Moore:1989vd,DiFrancesco:1997nk} for standard reviews of RCFTs, and \cite{Frohlich:2006ch,Chang:2018iay,Bhardwaj:2017xup} for detailed discussions of non-invertible symmetries in 1+1 dimensions.

\subsubsection{Review of RCFT and Verlinde lines}

In an RCFT, there are finitely many representations (or modules) with respect to the chiral algebra, which can be the Virasoro algebra for the minimal models, or extended chiral algebras such as the Kac-Moody current algebra in the case of a WZW model. 
We label these representations of the extended chiral algebra by $a,b,c,\cdots$. 
The most fundamental data of an RCFT is the fusion rule between representations:
\ie\label{fusion}
a\times b= \sum_c N^c_{ab} c\,,~~~N^c_{ab}\in \mathbb{Z}_{\ge 0}\,.
\fe
This means that the tensor product of representations $a$ and $b$ contains those representations $c$ with nonzero $N^c_{ab}$. 
For every $a$, there is a unique conjugate (or dual) representation $a^*$ such that $N^1_{ab} =N^1_{ba}= \delta_{ab^*}$ \cite{Moore:1988qv,Kitaev:2005hzj}.\footnote{The chiral algebra representations we encountered in Sections \ref{sec:Ising} and \ref{sec:Fib} are all self-dual, i.e., $a=a^*$.} Here 1 stands for the identity (or vacuum) representation, which obeys $N^a_{b1}=N^a_{1b}=\delta_{ab}$. 
In particular, $1=1^*$ and $(a^*)^*=a$.

The local primary operators of the diagonal RCFT are obtained by pairing the holomorphic and antiholomorphic representations together in a diagonal way. 
The operators are denoted by $\phi_{a,a^*}(z,\bar z)$, and their conformal weights are denoted by $(h,\bar h) = (h_a , h_{a^*})= (h_a,h_a)$.
Via the operator-state correspondence, the local operator $\phi_{a, a^*}(z, \bar z)$ corresponds to a highest-weight state in a product of holomorphic and antiholomorphic representations denoted by $\H_a \otimes \overline{\H}_{a^*}$.
This allows us to write the Hilbert space on a circle as a diagonal direct sum:
\ie
{\cal H} = \bigoplus_a {\cal H}_a \otimes \overline{\cal H}_{a^*}\,.
\fe
We will refer to ${\cal H}_1 \otimes \overline{\cal H}_1$ and the corresponding local operators as the ``identity sector,'' which consists of the identity operator and its chiral descendants.

In a diagonal RCFT, the topological lines that commute with the chiral algebra (which includes the stress-energy tensor) are called the Verlinde lines, and are in one-to-one correspondence with the local primary operators \cite{Verlinde:1988sn,Petkova:2000ip,Chang:2018iay}. 
When inserted at a fixed time, they give rise to conserved operators that implement a global symmetry on Hilbert space. 
We denote the topological line corresponding to representation $a$ by ${\cal L}_a$.
It was shown in \cite{Verlinde:1988sn,Petkova:2000ip} that these lines obey the same fusion rule as the primary operators:
\ie\label{Lfusion}
{\cal L}_a \times {\cal L}_b = \sum_c N_{ab}^c {\cal L}_c\,.
\fe
${\cal L}_a$ is called invertible if we have  ${\cal L}_a \times {\cal L}_{a^*}={\cal L}_{a^*} \times {\cal L}_{a}=1$; otherwise, it is called non-invertible. 
Mathematically, these Verlinde lines are described by a fusion category $\cal F$, which generalizes the ordinary description of finite symmetries in terms of  groups and their 't Hooft anomalies \cite{Bhardwaj:2017xup,Chang:2018iay,Lin:2019kpn}.
The Verlinde lines  $\cal F$ are a particularly simple class of finite, internal, non-invertible global symmetries  in RCFT. 

Since the Verlinde lines commute with the chiral algebra, the action of ${\cal L}_a$ on the Hilbert space is entirely determined by the action on the primaries.
The action of Verlinde lines on primaries was defined in \cite{Verlinde:1988sn} as
\ie\label{Laction}
{\cal L}_a \ket{b} = \frac{S_{ab}}{S_{1b}}\ket{b}\,,
\fe
where $S_{ab}$ is the modular S-matrix.
This matrix is unitary and obeys $S_{ab}=S_{ba}= S_{a^*b}^*$; for example, the S-matrices for the Ising and Fibonacci fusion categories discussed in Sections \ref{sec:Ising} and \ref{sec:Fib} are
\ie
S^\text{Ising} =
\frac12\left(\begin{array}{ccc}1 & 1 &\sqrt{2}\\
1 & 1&-\sqrt{2} \\
\sqrt{2}& -\sqrt{2} &0
\end{array}\right)
\,,~~~~S^\text{Fib} = \frac{1}{\sqrt{2+\varphi}} 
\left(\begin{array}{cc}1 & \varphi \\\varphi & -1\end{array}\right)\,,
\fe
where the representations are ordered as $a=1,\epsilon,\sigma$ and $a=1,\Phi$, respectively. 
The definition \eqref{Laction} obeys the fusion rules \eqref{Lfusion} thanks to the Verlinde formula \cite{Verlinde:1988sn, Moore:1988uz}
\begin{equation}
\label{eqn:Verlinde_Formula}
    N_{ab}^{c} = \sum_{d} \frac{S_{ad} S_{bd} S^*_{d c}}{S_{1d}}.
\end{equation}

The eigenvalue of ${\cal L}_a$ on the vacuum state $\ket{1}$ is known as the quantum dimension $\langle {\cal L}_a\rangle = {S_{a1}/S_{11}}$, which is positive in unitary theories. It is 1 if ${\cal L}_a$ is invertible, and greater than 1 if ${\cal L}_a$ is non-invertible.\footnote{To see this, we first note that $\langle {\cal L}_{a^*}\rangle = S_{a^*1}/S_{11}= S_{a1}/S_{11}=\langle {\cal L}_a\rangle$. Second, Verlinde's formula implies $\langle {\cal L}_a\rangle \langle {\cal L}_b\rangle = \sum_c N^c_{ab}\langle{\cal L}_c\rangle$. It follows that $(\langle {\cal L}_a\rangle)^2=\langle {\cal L}_a\rangle\langle {\cal L}_{a^*}\rangle=1+\sum_{c\neq1} N^c_{aa^*}\langle {\cal L}_c\rangle$, and hence the proof.}
By the state-operator correspondence, a local primary operator $\phi_{b,b^*}$ commutes with ${\cal L}_a$ if $S_{ab}=S_{a1}$, i.e., if ${\cal L}_a\ket{b} = \langle{\cal L}_a\rangle\ket{b}$:
\ie\label{commute}
{\cal L}_a \,\phi_{b,b^*} = \phi_{b,b^*} \,{\cal L}_a ~\Leftrightarrow~
 \frac{S_{ab} S_{11}}{S_{1a} S_{1b}}=1\,.
\fe
Since $S_{ab}$ is non-degenerate, it follows that  only the identity operator $\phi_{0,0}=1$ commutes with the entire fusion category $\cal F$. 
  
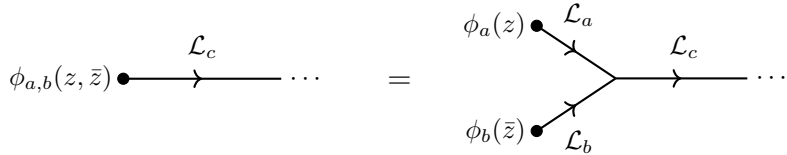
\begin{figure}[h!]
\centering
\raisebox{.8cm}{
\begin{tikzpicture}[scale=0.7]

  \filldraw[black] (-3,0) circle (3pt) node[anchor=east]{\footnotesize$\phi_{a,b}(z,\bar z)$};

\draw[thick, black, postaction={decorate},
    decoration={markings, mark=at position 0.5 with {\arrow{>}}}] (-3,0) --( 0,0);
     
    \node at (-1.5, .6) {\footnotesize \textbf{${\cal L}_c$}};
    
    \node at (.5, 0) {\footnotesize \textbf{$\cdots$}};
\end{tikzpicture}
}
\raisebox{1cm}{~~~$=$~~~}
\begin{tikzpicture}[scale=0.7]

  \filldraw[black] (-3,1) circle (3pt) node[anchor=east]{\footnotesize$\phi_{a}(z)$};
    \filldraw[black] (-3,-1) circle (3pt) node[anchor=east]{\footnotesize$\phi_{b}(\bar z)$};

\draw[thick, black, postaction={decorate},
    decoration={markings, mark=at position 0.5 with {\arrow{>}}}] (-1.5,0) --( 1,0);

    \draw[thick, black, postaction={decorate},
    decoration={markings, mark=at position 0.5 with {\arrow{>}}}] (-3,1) --( -1.5,0);
      \draw[thick, black, postaction={decorate},
    decoration={markings, mark=at position 0.5 with {\arrow{>}}}] (-3,-1) --( -1.5,0);
     
    \node at (-.2, .6) {\footnotesize \textbf{${\cal L}_c$}};
     \node at (-2.2, 1.2) {\footnotesize \textbf{${\cal L}_a$}};
    \node at (-2.2, -1.2) {\footnotesize \textbf{${\cal L}_b$}};
    \node at (1.5, 0) {\footnotesize \textbf{$\cdots$}};
\end{tikzpicture}
\caption{A twisted sector operator $\phi_{a,b}(z,\bar z)$  attached to a Verlinde line ${\cal L}_c$ can be factorized into a holomorphic and an antiholomorphic operators $\phi_a(z)$ 
 and $\phi_b(\bar z)$ connected by a network of lines. }\label{fig:Nabc}
\end{figure}

Given a Verlinde line ${\cal L}_c$, we consider the set of  point operators that can live at the end of ${\cal L}_c$. 
They are known as the twisted primaries,  defect operators, or disorder operators. 
A twisted primary is labeled by a holomorphic representation $a$ and an antiholomorphic representation $b$.
We denote a twisted primary by $\phi_{a,b}(z,\bar z)$, with conformal weight $(h_a, h_b)$.
Note that unlike local primary operators, generic twisted primaries in a diagonal RCFT have unequal left and right conformal weights. 
Given the line ${\cal L}_c$, the set of $\phi_{a,b}$'s  that reside at its endpoints consists of those satisfying the condition \cite{Petkova:2000ip}:
\ie \label{eq:disorder-condition}
N_{ab}^c\neq0 \,.
\fe
This condition can also be understood by holomorphically factorizing $\phi_{a,b}(z,\bar z)$ as in Figure \ref{fig:Nabc}.

Again, it is a misnomer to refer to $\phi_{a,b}$ as a  twisted sector ``operator" since it is not an operator acting on the Hilbert space of the RCFT. 
What does act on a Hilbert space is a pair of twisted primaries, $\phi_{a,b}(z_1,\bar z_1)$ and $\phi_{a',b'}(z_2,\bar z_2)$, connected by a Verlinde line ${\cal L}_c$:
\ie\label{pair}
\phi_{a,b} \overset{~~~~{\cal L}_c~~~~}{\xrightarrow{\hspace*{1cm}}}\phi_{a',b'} \,.
\fe
From equation \eqref{eq:disorder-condition}, we see that this pair of operators exists in the RCFT if $N_{ab}^c N_{a'b'}^{c^*}\neq0$.
When ${\cal L}_c=1$ is the identity line, this is a pair of local operators, and we have $b=a^*, b'= (a')^*$.

\subsubsection{Bulk perspective}

Here we discuss a 2+1d bulk perspective of RCFTs and their symmetries, which will become useful for later arguments. 
Via the standard Chern-Simons/WZW correspondence, one associates a 2+1d TQFT $\cal C$, which is typically a Chern-Simons gauge theory, to every 1+1d RCFT \cite{Moore:1988uz,Moore:1988qv,Witten:1988hf,Moore:1989yh,Elitzur:1989nr}. 
See also \cite{Komargodski:2020mxz,Chang:2020imq,Lin:2022dhv} for more recent discussions. 
The anyons (or Wilson lines) of the TQFT are denoted by $a,b,c,\cdots$. 
The fusion rule of these anyons is given by \eqref{fusion}. 
 Given an anyon $a$, there is a unique dual anyon $a^*$ (i.e., the antiparticle), satisfying $a\times  a^*= a^* \times a=1+\cdots$, where $1$ denotes the trivial anyon and $\cdots$ stands for other nontrivial anyons.  
An anyon $a$ is called abelian if $a\times a^*=a^*\times a=1$, and the corresponding Verlinde line ${\cal L}_a$ is invertible.  It is called  non-abelian otherwise.\footnote{The term ``non-abelian anyon" can sometimes be misleading in certain contexts. The fusion rule for non-abelian anyons is always commutative, i.e., $N_{ab}^c=N^c_{ba}$ and $a\times b=b\times a$. However, they do not form an abelian group, but an algebra. It is better to refer to them as ``non-invertible anyons" as they generate non-invertible 1-form global symmetries (see, e.g., \cite{Roumpedakis:2022aik}).} 
Mathematically, this 2+1d TQFT is described by a  modular tensor category (MTC), which will also be denoted by $\cal C$, and the anyons are its  simple objects. 
The fusion category $\cal F$ describing the Verlinde lines is obtained from  $\cal C$ by forgetting the braiding structure. 
Interested readers are referred to \cite{Moore:1988qv,Kitaev:2005hzj,Barkeshli:2014cna} for more detailed discussions of MTC.

The TQFT $\cal C$ is called abelian if every anyon is abelian. In this case the fusion category $\cal F$ generated by the Verlinde lines is a finite, invertible group-like symmetry. 
The TQFT $\cal C$ is called non-abelian if it contains at least one non-abelian anyon. In this case, the fusion category $\cal F$   is non-invertible.

One important piece of MTC data that we will use is the linking between two anyons $a,b$.
When we unlink two anyon loops $a$ and $b$, we pick up a coefficient $S_{ab}S_{11}/(S_{1a}S_{1b})$: 
\begin{equation}
\tikz[baseline=-0.5ex] 
    {
      \draw[->- = .6 rotate 10, thick, teal] (-1,0.2) arc[start angle=380, end angle=40, radius=0.4cm]; 
      \draw[->- = .6 rotate 10, thick, red] (-1.1,-0.1) arc[start angle=560, end angle=225, radius=0.4cm];
      \node[below] at (-0.3,-0.2) {\textcolor{red}{$b$}};
      \node[below] at (-1.8,-0.2) {\textcolor{teal}{$a$}};
    }
  =\frac{S_{ab}S_{11}}{S_{1a}S_{1b}}
  \tikz[baseline=-0.5ex]\node[below] at (0,0) {\textcolor{teal}{$a$}};
    \tikz[baseline=-0.5ex] \draw[->- = 0.52 rotate 10, thick, teal, thick, teal] (0,0) arc[start angle=360, end angle=0, radius=0.4cm];
    \tikz[baseline=-0.5ex]\node[below] at (0.5,0) {\textcolor{red}{$b$}};
    \tikz[baseline=-0.5ex] \draw[->- = 0.52 rotate 10, thick, red, thick, red] (0.5,0) arc[start angle=360, end angle=0, radius=0.4cm];
\end{equation}
When  $a$ and $b$ are both abelian, this is a phase factor associated with the braiding of the anyons. 
The condition of modularity in an MTC means that every anyon has nontrivial linking with at least one anyon (which can be itself) in the TQFT.
This is equivalent to requiring $S_{ab}$ to be unitary \cite{Kitaev:2005hzj}.

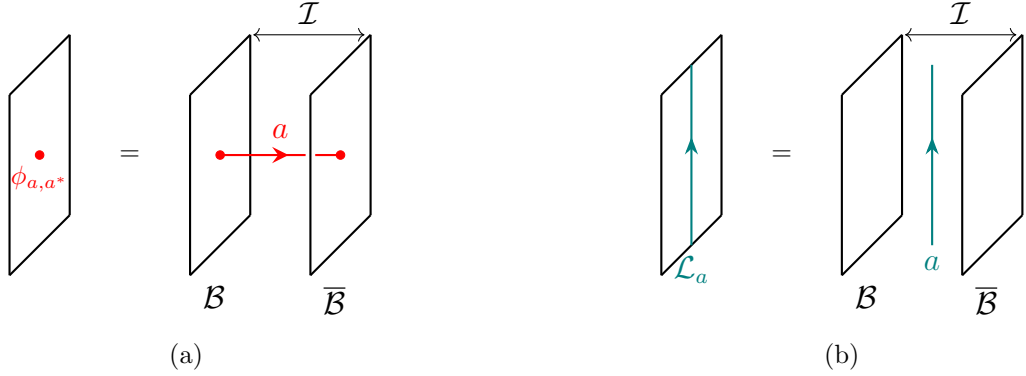
\begin{figure}[t]
    \centering
    \subcaptionbox{}[0.49\linewidth]
    {%
  
  \begin{tikzpicture}[scale = .8]
       
     \draw[thick](-1-3,0)--(-1-3,-3);
    \draw[thick](-1-3,0)--(0-3,1);
    \draw[thick](0-3,1)--(0-3,-2);
    \draw[thick](-1-3,-3)--(0-3,-2);    \filldraw[red] (-.5-3,-1) circle (2pt) ;
    \node[below,red] at (-3.5,-1) {\footnotesize$\phi_{a,a^*}$};
    \node[black] at (-2,-1) {\footnotesize$=$};

        \draw[thick](-1,0)--(-1,-3);
        \draw[thick](-1,0)--(0,1);
        \draw[thick](0,1)--(0,-2);
        \draw[thick](-1,-3)--(0,-2);
        \filldraw[red] (-0.5,-1) circle (2pt);
        \filldraw[red] (1.5,-1) circle (2pt);
        \draw[thick, red,  ->- = .8 rotate 0](-0.5,-1)--(0.92,-1);
        \draw[thick, red](1.08,-1)--(1.5,-1);
        \draw[<->] (0.05,1) -- (1.9,1) node[midway,above] {$\mathcal{I}$};
        %
        \draw[thick](1,0)--(1,-3);
        \draw[thick](1,0)--(2,1);
        \draw[thick](2,1)--(2,-2);
        \draw[thick](1,-3)--(2,-2);
        \node[below] at (-0.6,-3) {$\mathcal{B}$};
        \node[below] at (1.4,-3) {$\overline{\mathcal{B}}$};
        \node[above] at (0.5,-0.9) {\textcolor{red}{$a$}};

    \end{tikzpicture}
    }
     \subcaptionbox{}[0.49 \linewidth]
    {
    
    \begin{tikzpicture}[scale = .8]

     \draw[thick](-1-3,0)--(-1-3,-3);
    \draw[thick](-1-3,0)--(0-3,1);
    \draw[thick](0-3,1)--(0-3,-2);
    \draw[thick](-1-3,-3)--(0-3,-2);      \draw[thick, teal, postaction={decorate},
    decoration={markings, mark=at position 0.6 with {\arrow{Stealth}}}](-0.5-3,-2.5) -- (-0.5-3,0.5);
    \node[below] at (-0.5-3, -2.5) {\textcolor{teal}{${\cal L}_a$}};

    \node[black] at (-2,-1) {\footnotesize$=$};
    
        \draw[thick](-1,0)--(-1,-3);
        \draw[thick](-1,0)--(0,1);
        \draw[thick](0,1)--(0,-2);
        \draw[thick](-1,-3)--(0,-2);
        \draw[thick, teal, postaction={decorate},
    decoration={markings, mark=at position 0.6 with {\arrow{Stealth}}}](0.5, -2.5) -- (0.5, 0.5);
        \node[below] at (0.5, -2.5) {\textcolor{teal}{$a$}};
       
        \draw[<->] (0.05,1) -- (1.9,1) node[midway,above] {$\mathcal{I}$};
       
        \draw[thick](1,0)--(1,-3);
        \draw[thick](1,0)--(2,1);
        \draw[thick](2,1)--(2,-2);
        \draw[thick](1,-3)--(2,-2);
        \node[below] at (-0.6,-3) {$\mathcal{B}$};
        \node[below] at (1.4,-3) {$\overline{\mathcal{B}}$};
      
    \end{tikzpicture}
    }
    \caption{(a) A local primary operator $\phi_{a,a^*}(z,\bar z)$ is realized as an anyon line $a$ stretched between the two boundaries in the 2+1d TQFT $\cal C$. (b) The  Verlinde line ${\cal L}_a$ in the RCFT is identified with an anyon line $a$ in the 2+1d TQFT parallel to the boundary. }
    \label{fig:RCFT_TQFT}
\end{figure}

Physically, the 2+1d TQFT is related to the RCFT as follows. We start with the TQFT on an interval $\mathcal{I}$, and impose a gapless, holomorphic boundary condition $\cal B$ on one end of the interval, and a gapless, anti-holomorphic boundary condition $\overline{\cal B}$ on the other end \cite{Elitzur:1989nr,Fuchs:2002cm,Kapustin:2010if}, as shown in Figure \ref{fig:RCFT_TQFT}.
Since the bulk theory is topological (but the boundary conditions are not), one can shrink the length of the interval to obtain a 1+1d theory, which is the RCFT. 
From this bulk perspective, the local primary operator $\phi_{a,a^*}$ of the RCFT is an anyon line $a$ streched between the two boundaries. On the other hand, the Verlinde line ${\cal L}_a$ is mapped to an anyon line $a$ parallel to the boundary.
This gives a physical explanation for why the local primaries and the Verlinde lines are both labeled by the anyons. 
Furthermore, the action of  a Verlinde line ${\cal L}_a$  on a local primary operator $\phi_{b,b^*}$ in \eqref{Laction} is equivalent to the linking between two anyons $a,b$ (see, e.g., \cite{Komargodski:2020mxz}). 
In particular, ${\cal L}_a$
commutes with   $\phi_{b,b^*}$   if and only if  the linking between  $a,b$ is trivial, i.e., $S_{ab}S_{11}/(S_{1a}S_{1b})=1$.

\subsubsection{Symmetric sector}

Now we restrict ourselves to the set of operators in the RCFT that commute with all Verlinde lines ${\cal L}_a$ in $\cal F$. 
This $\cal F$-symmetric sector is generated by the local operators in the identity sector ${\cal H}_1\otimes \overline{\cal H}_1$, which consists of the identity operator and its chiral descendants.  

To see if the identity sector violates additivity or Haag duality, we need to determine which pairs of twisted primaries \eqref{pair} commute with the full fusion category $\cal F.$
As in section \ref{sec:Ising}, we will find violations of additivity when the identity sector contains a bilocal operator (which we can think of as \eqref{pair} with a trivial line).
As in section \ref{sec:Fib}, we will find violations of Haag duality when the identity sector contains a pair of twisted primaries connected by a nontrivial line that commutes with all bilocal operators in this sector.
There are several ways to determine which pairs of twisted primaries commute with $\cal F$; we will take a shortcut to the final answer by using the OPE trick from Section \ref{subsec:symmetric-sectors}.

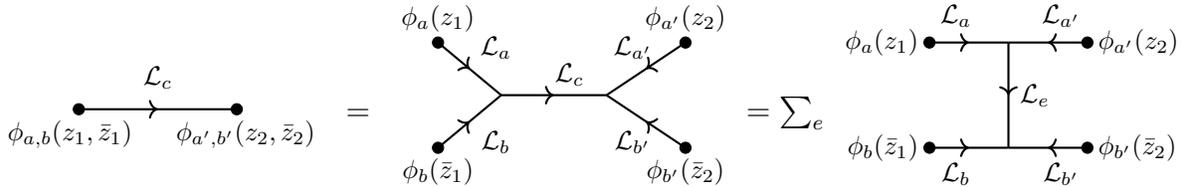
\begin{figure}[h!]
\centering
\raisebox{.5cm}{
\begin{tikzpicture}[scale=0.7]

  \filldraw[black] (-3,0) circle (3pt) node[anchor=north]{\footnotesize$\phi_{a,b}(z_1,\bar z_1)~~$};

  \filldraw[black] (0,0) circle (3pt) node[anchor=north]{\footnotesize$~~\phi_{a',b'}(z_2,\bar z_2)$};

\draw[thick, black, postaction={decorate},
    decoration={markings, mark=at position 0.5 with {\arrow{>}}}] (-3,0) --( 0,0);
     
    \node at (-1.5, .6) {\footnotesize \textbf{${\cal L}_c$}};
\end{tikzpicture}
}
\raisebox{1cm}{$=$~}
\begin{tikzpicture}[scale=0.7]

  \filldraw[black] (-3,1) circle (3pt) node[anchor=south]{\footnotesize$\phi_{a}(z_1)$};
    \filldraw[black] (-3,-1) circle (3pt) node[anchor=north]{\footnotesize$\phi_{b}(\bar z_1)$};
  \filldraw[black] (1.7,1) circle (3pt) node[anchor=south]{\footnotesize$\phi_{a'}(z_2)$};
    \filldraw[black] (1.7,-1) circle (3pt) node[anchor=north]{\footnotesize$\phi_{b'}(\bar z_2)$};

\draw[thick, black, postaction={decorate},
    decoration={markings, mark=at position 0.5 with {\arrow{>}}}] (-1.8,0) --( .2,0);

    \draw[thick, black, postaction={decorate},
    decoration={markings, mark=at position 0.5 with {\arrow{>}}}] (-3,1) --( -1.8,0);
      \draw[thick, black, postaction={decorate},
    decoration={markings, mark=at position 0.5 with {\arrow{>}}}] (-3,-1) --( -1.8,0);

        \draw[thick, black, postaction={decorate},
    decoration={markings, mark=at position 0.5 with {\arrow{>}}}] (1.7,1) --( .2,0);
      \draw[thick, black, postaction={decorate},
    decoration={markings, mark=at position 0.5 with {\arrow{>}}}] (1.7,-1) --(.2,0);
     
    \node at (-.5, .4) {\footnotesize \textbf{${\cal L}_c$}};
     \node at (-1.9, .9) {\footnotesize \textbf{${\cal L}_a$}};
    \node at (-1.9,- .9) {\footnotesize \textbf{${\cal L}_b$}};
    \node at (.7, .9) {\footnotesize \textbf{${\cal L}_{a'}$}};
    \node at (.7, -.9) {\footnotesize \textbf{${\cal L}_{b'}$}};
\end{tikzpicture}
\raisebox{1cm}{$=\sum_e$}
\begin{tikzpicture}[scale=0.7]

  \filldraw[black] (-1.5,1) circle (3pt) node[anchor=east]{\footnotesize$\phi_{a}(z_1)$};
    \filldraw[black] (-1.5,-1) circle (3pt) node[anchor=east]{\footnotesize$\phi_{b}(\bar z_1)$};
  \filldraw[black] (1.5,1) circle (3pt) node[anchor=west]{\footnotesize$\phi_{a'}(z_2)$};
    \filldraw[black] (1.5,-1) circle (3pt) node[anchor=west]{\footnotesize$\phi_{b'}(\bar z_2)$};

\draw[thick, black, postaction={decorate},
    decoration={markings, mark=at position 0.5 with {\arrow{>}}}] (-1.5,1) --(0,1);
\draw[thick, black, postaction={decorate},
    decoration={markings, mark=at position 0.5 with {\arrow{>}}}] (1.5,1) --( 0,1);
\draw[thick, black, postaction={decorate},
    decoration={markings, mark=at position 0.5 with {\arrow{>}}}] (-1.5,-1) --(0,-1);
\draw[thick, black, postaction={decorate},
    decoration={markings, mark=at position 0.5 with {\arrow{>}}}] (1.5,-1) --( 0,-1);
\draw[thick, black, postaction={decorate},
    decoration={markings, mark=at position 0.5 with {\arrow{>}}}] (0,1) --( 0,-1);

    \node at (.5, 0) {\footnotesize \textbf{${\cal L}_e$}};
     \node at (-1, 1.5) {\footnotesize \textbf{${\cal L}_a$}};
    \node at (-1, -1.5) {\footnotesize \textbf{${\cal L}_b$}};
    \node at (1, 1.5) {\footnotesize \textbf{${\cal L}_{a'}$}};
    \node at (1, -1.5) {\footnotesize \textbf{${\cal L}_{b'}$}};
\end{tikzpicture}
\caption{By performing a crossing move for the Verlinde lines, the OPE can be performed holomorphically. (Here we have suppressed the coefficients, known as the F-symbols, in the crossing move.)}\label{fig:crossing}
\end{figure}

The OPE limit of the twisted primaries in equation \eqref{pair} gives a sum of local primary operators.
This OPE can be performed by first holomorphically factorizing $\phi_{a,b}$ and $\phi_{a',b'}$, and then performing a crossing move on the Verlinde lines as in Figure \ref{fig:crossing}. 
It follows that the local primary operator $\phi_{e,e^*}$ appears in the $\phi_{a,b} \overset{~~~~{\cal L}_c~~~~}{\xrightarrow{\hspace*{1cm}}}\phi_{a', b'}$ OPE if and only if we have
\ie
N^e_{aa'}N^{e^*}_{bb'} \neq0 \,.
\fe

For \eqref{pair} to commute with all Verlinde lines, the OPE must contain only the identity operator, since this is the only primary that commutes with every symmetry in $\cal F.$
From this observation, we see that the conditions for \eqref{pair} to be ${\cal F}$-symmetric are
\begin{equation}
    N^1_{aa'}=N^1_{bb'}=1
\end{equation}
and
\begin{equation}
    N^e_{aa'}N^{e^*}_{bb'} =0 \quad \forall ~ e \neq 1.
\end{equation}
The first condition in particular implies $a'=a^*,b'=b^*$. 
So the most general pair of twisted primaries that commute with $\cal F$ is of the form
\ie\label{Csym}
&\phi_{a,b} \overset{~~~~{\cal L}_c~~~~}{\xrightarrow{\hspace*{1cm}}}\phi_{a^*,b^*} \,,\\
&N^c_{ab}\neq0\,,~~ N^e_{aa^*}N^{e^*}_{bb^*} =0\,,~~~\forall ~e\neq 1\,.
\fe
In other words, the condition is that $a\times a^*$ and $b\times b^*$ must share no nontrivial common channel. 
Such operators are referred to as the ``patch  operators" in \cite{Ji:2019jhk,Wu:2020yxa,Chatterjee:2022kxb,Chatterjee:2022jll,Inamura:2023ldn}.  
In particular, because the only channel in the $1 \times 1$ OPE is trivial, any pair of holomorphic twisted primaries $\phi_{a,1} \overset{~~~~{\cal L}_a~~~~}{\xrightarrow{\hspace*{1cm}}}\phi_{a^*,1}$ will always commute with $\cal F$.

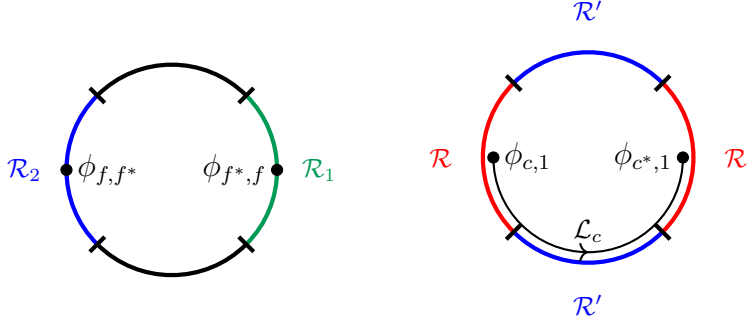
\begin{figure}[h!]
\centering
\raisebox{.5cm}{\begin{tikzpicture}[scale=0.7]  
    \draw[ultra thick] (45:2) arc[start angle=45, end angle=135, radius=2];

      \draw[ultra thick] (225:2) arc[start angle=225, end angle=315, radius=2];

    \draw[ultra thick, ForestGreen] (315:2) arc[start angle=-45, end angle=45, radius=2];

    \draw[ultra thick, blue] (135:2) arc[start angle=135, end angle=225, radius=2];

    \foreach \angle in { 45, 135,  225, 315} {
   
        \pgfmathsetmacro\x{2*cos(\angle)}
        \pgfmathsetmacro\y{2*sin(\angle)}
        
        \pgfmathsetmacro\dx{0.2*cos(\angle)}
        \pgfmathsetmacro\dy{0.2*sin(\angle)}
        
        \draw[ultra thick] (\x-\dx, \y-\dy) -- (\x+\dx, \y+\dy);
    }

\filldraw[black] (-2,0) circle (3pt) node[anchor=west]{$\phi_{f,f^*}$};

\filldraw[black] (2,0) circle (3pt) node[anchor=east]{$\phi_{f^*,f}$};

    \node at (2.8, 0) {\footnotesize \textbf{\color{ForestGreen}$\R_1$}};
    \node at (-2.8, 0) {\footnotesize \textbf{\color{blue}$\R_2$}};

\end{tikzpicture}}
~~~~~
\begin{tikzpicture}[scale=0.7]  
    \draw[ultra thick,blue] (45:2) arc[start angle=45, end angle=135, radius=2];

      \draw[ultra thick,blue] (225:2) arc[start angle=225, end angle=315, radius=2];

    \draw[ultra thick, red] (315:2) arc[start angle=-45, end angle=45, radius=2];

    \draw[ultra thick, red] (135:2) arc[start angle=135, end angle=225, radius=2];

    \draw[thick, black, postaction={decorate},
    decoration={markings, mark=at position 0.5 with {\arrow{>}}}] 
    (180:1.8) arc[start angle=180, end angle=360, radius=1.8];

    \foreach \angle in { 45,  135,  225,  315} {
        \pgfmathsetmacro\x{2*cos(\angle)}
        \pgfmathsetmacro\y{2*sin(\angle)}
        
  \pgfmathsetmacro\dx{0.2*cos(\angle)}  
        \pgfmathsetmacro\dy{0.2*sin(\angle)}
        \draw[ultra thick] (\x-\dx, \y-\dy) -- (\x+\dx, \y+\dy);
    }

  \filldraw[black] (-1.8,0) circle (3pt) node[anchor=west]{$\phi_{c,1}$};

\filldraw[black] (1.8,0) circle (3pt) node[anchor=east]{$\phi_{c^*,1}$};

   \node at (0, -1.4) {\footnotesize \textbf{${\cal L}_c$}};

    \node at (2.8, 0) {\footnotesize \textbf{\color{red}$\R$}};
    \node at (-2.8, 0) {\footnotesize \textbf{\color{red}$\R$}};
    \node at (0, 2.8) {\footnotesize \textbf{\color{blue}$\R'$}};
    \node at (0, -2.8) {\footnotesize \textbf{\color{blue}$\R'$}};
\end{tikzpicture}
\caption{Left: The identity sector of an RCFT violates additivity if the associated  2+1d TQFT $\cal C$ has an abelian anyon. Right: The identity sector violates Haag duality if there the  exists an anyon $c$ that has trivial linking with every abelian anyon. We show that the latter condition is equivalent to the existence of a non-abelian anyon in $\cal C$.}\label{fig:RCFT}
\end{figure}

As in section \ref{sec:Ising}, the identity sector of the RCFT violates additivity if there exists a pair of local operators that is $\cal F$-symmetric.
Applying the condition \eqref{Csym}, this requires a pair of local operators of the form $\phi_{f,f^*}$ and $\phi_{f^*,f}$ connected by  a trivial line (i.e., $c=1$), with $f$ satisfying
\ie
{\cal L}_f\times {\cal L}_{f^*} = {\cal L}_{f^*}\times{\cal L}_f=1 \,.
\fe
(See the left of Figure \ref{fig:RCFT}.)
In the RCFT language, this means that the Verlinde line ${\cal L}_f$ generates an ordinary, invertible, group-like symmetry. 
In the TQFT language, $f$ is an abelian anyon. 
We conclude that additivity is violated in the identity sector if $\cal F$ contains an invertible element.

To understand when Haag duality is violated, it will be convenient to use the 2+1d TQFT picture. 
 The identity sector of this RCFT violates  Haag duality if
 there is a pair of twisted primaries connected by a nontrivial line ${\cal L}_c$
 of the form \eqref{Csym}, such that it commutes with every bilocal operator in ${\cal R}'$ in the identity sector. 
 In particular, such an operator with $a=c,b=1$     always satisfies \eqref{Csym} (see the right of Figure \ref{fig:RCFT}). 
It commutes with every bilocal operator in the identity sector if
 \ie
\frac{S_{cf} S_{11}}{S_{1c}S_{1f}} =1\,,~~~~\forall~f\times f^*= f^*\times f =1\,.
 \fe 
 In the TQFT $\cal C$ language, this means that the linking between $c$ and every abelian anyon $f$ is trivial. 
 So a sufficient  condition for Haag duality violation is the existence of an anyon $c$ in  that links trivially with every abelian anyon in the corresponding TQFT $\cal C$.

The family of abelian anyons forms a subcategory $\mathcal{C}_{\text{ab}} \subseteq \mathcal{C}.$
The set of anyons that link trivially with every anyon in $\mathcal{C}_{\text{ab}}$ is called the \textit{centralizer} of this subcategory.
So in mathematical language, our sufficient condition for Haag duality violation is that $\mathcal{C}_{\text{ab}}$ should have a nontrivial centralizer.
If $\mathcal{C}$ is abelian, then we have $\mathcal{C}_{\text{ab}} = \mathcal{C},$ and non-degeneracy of $S_{ab}$ implies that the centralizer of $\mathcal{C}_{\text{ab}}$ is trivial.
So while we automatically have additivity violation in the abelian case, we do not find a violation of Haag duality using our techniques.
On the other hand, if we have $\mathcal{C}_{\text{ab}} \subsetneq \mathcal{C},$ then the centralizer is automatically nontrivial, because \textit{any} proper subcategory of a modular tensor category has a nontrivial centralizer --- this is implied, for example, by theorem 3.2(ii) of \cite{muger}.\footnote{Applying that theorem to the case of a \textit{modular} subcategory leads to the statement that any modular tensor category factorizes between a modular subcategory and its centralizer; this is a key result from \cite{muger}, and may be more familiar to experts than the intermediary theorem 3.2(ii).}
Consequently, we may conclude that whenever $\mathcal{C}$ contains a non-invertible element --- equivalently, whenever there is a non-abelian anyon --- Haag duality is violated in the symmetric sector.

Putting together our results, we have shown the following:
\textit{Let ${\cal H}_1\otimes \overline{\cal H}_1$ be the identity sector of a diagonal RCFT and $\cal F$ be the fusion category for the Verlinde lines. Then ${\cal H}_1\otimes \overline{\cal H}_1$ violates additivity if $\cal F$ contains a (non-trivial) invertible element, and violates Haag duality if $\cal F$ contains a  non-invertible element.} 
Equivalently, additivity is violated if the associated 2+1d TQFT $\cal C$ has an abelian anyon, and Haag duality is violated if $\cal C$ has a non-abelian anyon.

One immediate consequence is that whenever $\cal C$ is nontrivial, i.e., whenever the RCFT has more than just the identity sector, the latter must violates additivity or/and Haag duality. 
Our result is consistent with \cite{Benedetti:2024dku}, where it was shown that additivity and Haag duality together imply modular invariance. 
Indeed, the identity sector is  not modular invariant whenever $\cal C$ is nontrivial. 
Above we have provided an explicit condition for when additivity and Haag duality  are violated. 
When $\cal C$ is trivial, the RCFT is a tensor product of a holomorphic CFT and its antiholomorphic counterpart, such as the Monster$\times \overline{\text{Monster}}$ CFT. 
In this case, the identity sector is the entire CFT, which is modular invariant.

\begin{table}[h!]
\begin{center}
\begin{tabular}{|c|c|c|}
\hline 
Additivity & Haag duality & Examples \\
\hline
$\times$ & $\checkmark$ &  $\mathfrak{su}(n)_1, (\mathfrak{e}_6)_1,(\mathfrak{e}_7)_1$   \\
\hline
$\checkmark$ & $\times$ & $(\mathfrak{g}_2)_1,(\mathfrak{f}_4)_1$,  $(\mathfrak{g}_2)_2$ \\
\hline
$\times $ & $\times$ &  Ising, $\mathfrak{su}(2)_2,(\mathfrak{e}_8)_2$ \\
\hline
$\checkmark$ & $\checkmark$ & 
$(\mathfrak{e}_8)_1, |\text{Monster}|^2$\\
\hline 
\end{tabular}
\end{center}
\caption{Violation of additivity or/and Haag duality in the identity sector  of various diagonal RCFTs. The last row corresponds to holomorphically factorized CFTs whose identity sectors form full, modular invariant CFTs.}
\end{table}

\section{Lattice examples}
\label{sec:lattice}

In the previous section, we showed that in diagonal RCFTs, restricting to the symmetric sector of the Verlinde symmetry with both invertible and non-invertible elements results in violations of both additivity and Haag duality.
This section discusses two examples of non-invertible symmetries in 1+1D lattice systems, in which similar phenomena occur, though not always in exactly the same way as in the continuum.

Subsection \ref{sec:KW-lattice} discusses the Kramers-Wannier symmetry on a 1+1D lattice system; unlike in the continuum, the Kramers-Wannier symmetry on the lattice mixes with the spatial translation group in addition to the global $\mathbb{Z}_2$ symmetry \cite{Seiberg:2023cdc,Seiberg:2024gek}.
This leads to a violation of additivity and Haag duality in the symmetric sector that is qualitatively different from what was discussed in Section \ref{sec:Ising}.

Subsection \ref{sec:D8} discusses the second simplest example of  non-invertible symmetries on spin chains, the Rep(D$_8$) fusion category, in which the violation of additivity and Haag duality is exactly like those discussed in Section \ref{sec:continuum}.

\subsection{Kramers-Wannier operator in the Ising lattice model}
\label{sec:KW-lattice}

We consider a one-dimensional spatial lattice with $L$ links labeled by $\ell$.
We place a qubit on every link, and impose periodic boundary conditions by identifying $\ell\sim \ell+L$. 
The full operator algebra is the matrix algebra $\text{Mat}(2^L, \mathbb{C})$, as in equation \eqref{matrix}.

The simplest non-invertible symmetry on a one-dimensional lattice is the Kramers-Wannier operator \cite{Seiberg:2023cdc,Seiberg:2024gek}:
\ie
\mathsf{KW} = \sqrt{2} e^{ - \frac{2\pi i L}{8} }\prod_{\ell=1}^{L-1} \left(e^{ \frac{ i \pi}{4} X_\ell} e^{\frac{i \pi}{4} Z_\ell Z_{\ell+1}}\right)
e^{\frac{i\pi}{4} X_L} \frac{1+U}{2}\,,
\fe
where $U=\prod_{\ell=1}^LX_j$. 
See also \cite{Aasen:2016dop,Li:2023ani,Chen:2023qst} for closely related maps and operators. 
The operator $\mathsf{KW}$ is non-invertible and annihilates every state that is $\mathbb{Z}_2$-odd, i.e., $\mathsf{KW} \ket{\phi}=0$ if $U\ket{\phi}=-\ket{\phi}$. 
It can also be expressed in terms of a Matrix Product Operator (MPO)   of bond dimension 2 \cite{Tantivasadakarn:2021vel,Seiberg:2024gek,Gorantla:2024ocs}:
\ie\label{eq:KW-MPO}
\mathsf{KW} = \text{Tr}\left( \, (\mathbb{KW}^1 )(\mathbb{KW}^2) \cdots 
(\mathbb{KW}^L)\,\right)\,,
\fe
where
\ie 
\mathbb{KW}^\ell  = 
 \left(\begin{array}{cc}\ket{0}\bra{+}_\ell & \ket{0}\bra{-}_\ell \\\ket{1}\bra{-}_\ell& \ket{1}\bra{+}_\ell \end{array}\right)\,
\fe
is a 2-by-2 matrix whose individual entries are operators acting on the $\ell$-th physical qubit. 
The trace is taken over the 2-dimensional bond space for the MPO matrix, rather than over the Hilbert space of physical qubits. Here $X\ket{\pm}= \pm \ket{\pm}, Z\ket{0}=\ket{0},Z\ket{1}=-\ket{1}$.

The Kramers-Wannier operator $\mathsf{KW}$ acts on the $\mathbb{Z}_2$-even local operators as
\ie
(\mathsf{KW}) X_\ell  =Z_\ell Z_{\ell+1} (\mathsf{KW})\,,~~~~
(\mathsf{KW}) Z_\ell Z_{\ell+1}  =X_{\ell+1} (\mathsf{KW})\,.
\fe
As an example,  the critical transverse-field Ising Hamiltonian commutes with the Kramers-Wannier operator:
\ie\label{TFIM}
H = - \sum_{\ell=1}^L X_\ell - \sum_{\ell=1}^L Z_\ell Z_{\ell+1}\,.
\fe
The thermodynamic limit $L\to\infty$ of \eqref{TFIM} is described by the Ising CFT, and in this limit the lattice symmetries $U$ and $\mathsf{KW}$ become the Verlinde lines ${\cal L}_\epsilon$ and ${\cal L}_\sigma$ from Section \ref{sec:Ising}.

Together, the non-invertible Kramers-Wannier operator $\mathsf{KW}$, the $\mathbb{Z}_2$ operator $U$, and lattice translation $T:\ell\to \ell+1$ form the following algebra \cite{Seiberg:2023cdc,Seiberg:2024gek}:
\ie \label{eq:KW-lattice}
&U^2=1\,,~~~T^L=1\,,~~~TU=UT\,,\\
&U (\mathsf{KW}) = (\mathsf{KW}) U = \mathsf{KW}\,,~~~(\mathsf{KW})^\dagger = T^{-1} (\mathsf{KW}) = (\mathsf{KW} )T^{-1}\,,\\
&(\mathsf{KW})^2 = (1+U)T\,.
\fe
Importantly, this algebra mixes the Kramers-Wannier symmetry with lattice translation. In the thermodynamic limit $L\to\infty$, the lattice translation becomes trivial on the low-lying states, i.e., $T\sim1$, and we recover the  the algebra \eqref{Isingfusion} in the continuum Ising CFT.

The operators symmetric under the full algebra \eqref{eq:KW-lattice} are
\ie
{\cal A}_\mathsf{KW} (S^1) &= \left\{ \, 
{\cal O}\in \text{Mat}(2^L,\mathbb{C})~\Big|~
(\mathsf{KW}){\cal O} = {\cal O} \,(\mathsf{KW}),\,
U{\cal O} = {\cal O} \,U ,\,
T{\cal O} = {\cal O} \,T
\,\right\}\\
&= \Big\langle\,
\sum_{\ell=1}^L (X_\ell +Z_\ell Z_{\ell+1})\,,~~
\sum_{\ell=1}^L (X_\ell Z_{\ell+1}Z_{\ell+2}+Z_\ell Z_{\ell+1}X_{\ell+2})
,\,\cdots \Big\rangle
\fe
Note that while a pair of $Z$ operators commutes with the $\mathbb{Z}_2$ operator $U$, it does not commute with $\mathsf{KW}$. In fact, 
 the Kramers-Wannier operator maps a pair of $Z$ order operators to a disorder operator $U(\ell_1,\ell_2) = \prod_{\ell=\ell_1}^{\ell_2} X_\ell$:
\ie\label{latticeKW}
(\mathsf{KW}) \, Z_{\ell_1-1}Z_{\ell_2} = X_{\ell_1}X_{\ell_1+1}\cdots X_{\ell_2}
(\mathsf{KW})\,.
\fe
This is the lattice counterpart of Figure \ref{fig:KW}. 

Since the lattice algebra includes the generator of translations, any nontrivial symmetric operator must have support on the full spin chain.
From this we deduce the identity
\ie\label{AKW}
{\cal A}_\mathsf{KW}(\R)=\{1\}\,~~~ \text{if}~~ \R\subsetneq S^1, 
\fe
which immediately leads to violations of both additivity and Haag duality. 
Additivity is violated by considering the circle $S^1$ (which has a nontrivial algebra) as a union of two overlapping intervals (which have trivial algebras).
Haag duality is violated because the commutant of an interval algebra contains all symmetric operators, but these are not contained in the algebra of the complementary interval. 
The additivity and Haag duality violations in this case are qualitatively different from the examples in section \ref{sec:continuum}, which concern \textit{internal} global symmetries that do not mix with spacetime symmetries. 
The lattice Kramers-Wannier symmetry is not internal since its algebra cannot be disentangled from the lattice translation. 
See \cite{Seifnashri:2023dpa,ParayilMana:2024txy,Chatterjee:2024ych,Lu:2024ytl,Cao:2024qjj,Pace:2024tgk,Gorantla:2024ocs,Seo:2024its,Ma:2024ypm,Pace:2024oys} for other interesting lattice symmetries that mix with translations and  more general crystalline symmetries.

 \subsection{Rep(D$_8$) fusion category}
 \label{sec:D8}

 Next we consider an internal lattice non-invertible symmetry that does not mix with lattice translations.
 Its symmetric subalgebra also violates both additivity and Haag duality.  
 Our setup is again a closed periodic chain with $L$ links and a qubit placed on each link.
 In the present example, we require $L$ to be even.
 
Consider the following operator \cite{Seifnashri:2025fgd}:
\ie
	\mathsf{D} &=  \Tr (\mathbb{D}^1 \mathbb{D}^2 \cdots \mathbb{D}^L) \,,~~~~
\mathbb{D}^\ell& =  \frac{1}{\sqrt2} \begin{pmatrix}
1 & X_\ell \\
1 & -X_\ell
\end{pmatrix}\,,
\fe
where we have used the MPO notation introduced in equation \eqref{eq:KW-MPO}. 
(See also\cite{Fechisin:2023dkj,Seifnashri:2024dsd,Choi:2024rjm,Inamura:2024jke,Pace:2024acq,Warman:2024lir,Cao:2025qnc} for different lattice constructions of this symmetry.)
We define $\mathbb{Z}_2$ operators $U^{\text{e}}$ and $U^{\text{o}}$ that act only on the even or odd sites respectively, given by
\ie
 U^\text{e} = \prod_{n=1}^{L/2}X_{2n}\,,~~~~U^\text{o} = \prod_{n=1}^{L/2}X_{2n-1}\,.
\fe
The symmetry subgroups generated by these operators are called $\mathbb{Z}_{2}^{\text{e}}$ and $\mathbb{Z}_{2}^{\text{o}},$ respectively.
The operators $\mathsf{D}, U^{\text{e}},$ and $U^{\text{o}}$ obey the following operator algebra:
\ie\label{repd8}
&U^\text{e} \,\mathsf{D}=U^\text{o}\,  \mathsf{D}=\mathsf{D}\, U^\text{e} =\mathsf{D}\,U^\text{o}  =\mathsf{D}\,,\\
&\mathsf{D}^2 = (1+U^\text{e})(1+U^\text{o})\,, \\
&\mathsf{D}^{\dagger} = \mathsf{D}.
\fe
This is the representation ring of the dihedral group D$_8$ of order 8. 
These symmetries can be furthermore shown to generate a Rep(D$_8$) fusion category \cite{Seifnashri:2025fgd}. 
From \eqref{repd8}, it is clear that the operator $\mathsf{D}$ annihilates every state that is $\mathbb{Z}_2^\text{e}$-odd or $\mathbb{Z}_2^\text{o}$-odd, i.e., $\mathsf{D}\ket{\phi}=0$ if $U^\text{e}\ket{\phi} =-\ket{\phi}$ or $U^\text{o}\ket{\phi} =-\ket{\phi}$. 
Hence $\mathsf{D}$ has a nontrivial kernel, and is non-invertible. 
A concrete example of a family of Rep(D$_8$)-symmetric Hamiltonians is  \cite{Seifnashri:2025fgd},
\ie
H=  h_0 \sum_{\ell=1}^LX_\ell
+h_1 \sum_{\ell=1}^L Z_{\ell-1} (1+X_\ell)Z_{\ell+1}\,,
\fe
which is known as the zigzag model or the dual  XXZ model \cite{10.21468/SciPostPhysCore.4.2.010,Eck:2023gic,PhysRevB.108.L100304,Cao:2025qnc}. 
The continuum limit 
of this family of Hamiltonians gives the $c=1$ orbifold CFTs.

This non-invertible operator $\mathsf{D}$ acts on Pauli operators as 
\ie
\mathsf{D} X_\ell  = X_\ell \mathsf{D} \,,~~~~~
\mathsf{D} Z_{\ell-1}Z_{\ell+1}  = Z_{\ell-1}X_\ell Z_{\ell+1}\mathsf{D}\,.
\fe
The Rep(D$_8$)-symmetric algebra is 
\ie\label{repd8bond}
&{\cal A}_\text{Rep(D$_8$)} (S^1)\equiv
\Big\{ 
{\cal O}\in \text{Mat}(2^L ,\mathbb{C}) ~\Big| ~\mathsf{D} {\cal O}= {\cal O} \mathsf{D} ,~{\cal O}U^\text{e} =U^\text{e} {\cal O}~,{\cal O}U^\text{o} =U^\text{o} {\cal O}
\Big\}\\
&= \Bigl\langle
X_\ell , Z_{\ell-1}(1+X_\ell)Z_{\ell+1} ,
Z_{\ell-2} (1+X_{\ell-1}X_{\ell+1})Z_{\ell+2}, Z_{\ell-3}(1+X_{\ell-2}X_\ell X_{\ell+2})Z_{\ell+3},\cdots
\Bigr\rangle_{\ell} \,,
\fe
where $\dots$ include similar terms with two $Z$'s further separated, multiplied by 1 plus a product of $X$'s.

Note that the operator $Z_{\ell-1} (1-X_\ell)Z_{\ell+1}$ anticommutes with $\mathsf{D}$. 
Thus, in this symmetric sector, additivity can be violated by considering a pair of such operators localized in two disjoint intervals, similar to Figure \ref{fig:additivity}.

To violate Haag duality, we consider the disorder operator $U(\ell_1,\ell_2)=\prod_{\ell=\ell_1}^{\ell_2}X_\ell$, which commutes with $\mathsf{D}$. 
Take $\R$ to be two disconnected intervals, with one component containing $\ell_1$ and the other containing $\ell_2$ as in Figure \ref{fig:Haag}. 
As is clear from \eqref{repd8bond}, $U(\ell_1,\ell_2)$ commutes with all the operators in ${\cal A}_\text{Rep(D$_8$)}(\R')$. 
But $U(\ell_1,\ell_2) \notin {\cal A}_\text{Rep(D$_8$)}(\R)$. Hence we find a violation of Haag duality. 
(Note that unlike the $\mathbb{Z}_2$-symmetric sector discussed in Section \ref{sec:invertible}, a pair of $Z_\ell$'s is \textit{not} in ${\cal A}_\text{Rep(D$_8$)}(\R')$, and cannot be used to avoid the violation.)

We conclude that the Rep(D$_8$)-symmetric subalgebra violates both additivity and Haag duality. 

\section{Conclusions and outlook}

  Although it may seem counterintuitive, 
restricting to a subsector of a theory can sometimes reveal important global aspects about the full theory. 
In this work, we related the local algebraic properties, additivity and Haag duality, of a symmetric sector to the global symmetry of the full theory. 
In diagonal RCFTs, we showed that the identity sector, which is the symmetric sector of the finite global symmetry generated by the Verlinde lines, violates Haag duality (additivity) if the symmetry algebra contains a non-invertible (invertible) element. 
We also analyzed other examples of finite, internal, global symmetries in the continuum and on the lattice and found the same conclusion.

However, our setup is still somewhat special. 
First,  the invertible symmetry groups studied  here are all abelian finite groups, and we have not discussed non-abelian or continuous groups.
Second,  Verlinde lines in diagonal RCFTs are not the most general possible non-invertible symmetries in 1+1D. In mathematical terms, they are special in that   the associated fusion category can always be lifted to a modular tensor category. 
Finally, 
 the primary operators in a diagonal RCFT form a special representation, known as the regular module category, of the fusion category, and we have not discussed other representations. (This is the analog of a regular representation for a finite group.) 
It would certainly be interesting to apply this analysis to more general invertible and non-invertible symmetries in QFTs and lattice models.

One future direction is to extend this analysis to systems with higher-form global symmetries. In a companion paper with Harlow \cite{paper2}, we introduce a weaker notion of additivity we call \textit{disjoint additivity} by requiring $\R_1\cap\R_2=\emptyset$ in the definition \eqref{additivity} to accommodate higher-form global symmetries.\footnote{In all our examples of additivity violation within the symmetric sector of an internal global symmetry, we also observe a violation of disjoint additivity.} 
Another direction is to study spacetime symmetries in continuum field theory and crystalline symmetries in lattice systems. 
The simplest example is to consider the lattice translation operator $T$, whose symmetric algebra is the same as \eqref{AKW}. 
This gives an example of an invertible \textit{spacetime} symmetry whose symmetric sector violates both additivity \textit{and} Haag duality. 
While the distinction between internal and spacetime symmetries is unambiguous in relativistic QFT thanks to the Coleman-Mandula theorem, it is less so on the lattice. 
For instance, the lattice Kramers-Wannier symmetry mixes with lattice translations, and more general lattice crystalline symmetries can mix with internal global symmetries in the continuum limit  (see, e.g., \cite{Cheng:2022sgb,Barkeshli:2025cjs}, for recent discussions).

  \section*{Acknowledgements}

We thank Arkya Chatterjee, Yichul Choi, Daniel Harlow, Corey Jones, Anton Kapustin,  Hong Liu, John McGreevy, Greg Moore, Sal Pace, Nathan Seiberg, Sahand Seifnashri, David Simmons-Duffin, and Xiao-Gang Wen for helpful discussions. 
We are especially grateful to Daniel Harlow for many stimulating discussions and collaboration on a companion project \cite{paper2}. 
SHS is  supported by the Simons Collaboration on Ultra-Quantum Matter, which
is a grant from the Simons Foundation (651444, SHS). 
JS is supported by the DOE Early Career Award DE-SC0021886,  the Packard Foundation Award in Quantum Black Holes and Quantum Computation, the DOE QuantISED DE-SC0020360 contract 578218, and by the Heising-Simons Foundation grant 2023-443. 
MS is supported by the U.S. Department of Energy, Office of Science, Office of High Energy Physics of U.S. Department of Energy under grant Contract Number  DE-SC0012567 (High Energy Theory research) and by the Massachusetts Institute of Technology.
SHS is grateful to the Institute for Advanced Study (IAS)  for its hospitality during the course of this work. 
Part of this work was completed during the Kavli Institute for Theoretical Physics (KITP) program ``Generalized Symmetries in Quantum Field Theory: High Energy Physics, Condensed Matter, and Quantum Gravity", which is supported in part by grant NSF PHY-2309135 to the  KITP. 
The authors of this paper were ordered alphabetically.

\bibliography{ref}

\bibliographystyle{JHEP}

\end{document}